\documentclass[10pt,conference]{IEEEtran}
\IEEEoverridecommandlockouts
\usepackage{cite}
\usepackage{amsmath,amssymb,amsfonts}
\usepackage{algorithmic}
\usepackage{graphicx}
\usepackage{textcomp}
\usepackage{xcolor}
\usepackage{comment}
\usepackage{enumitem}
\def\BibTeX{{\rm B\kern-.05em{\sc i\kern-.025em b}\kern-.08em
    T\kern-.1667em\lower.7ex\hbox{E}\kern-.125emX}}

\newcommand {\ie} {{\em i.e., }}
\newcommand {\eg} {{\em e.g., }}
\newcommand {\beq} {\begin{equation}}
\newcommand {\eeq} {\end{equation}}
\newcommand {\bequn} {\begin{equation*}}
\newcommand {\eequn} {\end{equation*}}
\newcommand {\bear} {\begin{eqnarray}}
\newcommand {\eear} {\end{eqnarray}}
\newcommand {\bearun} {\begin{eqnarray*}}
\newcommand {\eearun} {\end{eqnarray*}}
\newcommand {\fig}[1]{Fig.~\ref{#1}}

\newcommand {\Eqref}[1]{Eq.~(\ref{#1})}

\DeclareMathOperator*{\map}{mAP}
\DeclareMathOperator*{\cdf}{cdf}
\DeclareMathOperator*{\argmax}{arg\,max}

\usepackage{enumitem}
\usepackage{multirow}
\usepackage{subcaption}
\usepackage{hyperref}
\usepackage{makecell}
\usepackage{pifont}
\newcommand{\cmark}{\ding{51}}
\newcommand{\xmark}{\ding{55}}

\newcommand{\useColor}{clean}  
\usepackage{xstring}
\usepackage{soul,xcolor} 
\setstcolor{red}

\newcommand{\highlight}[2]{%
    \IfEqCase{#1}{%
        {color}{\textcolor{blue}{#2}}%
        {clean}{#2}%
    }[\PackageError{highlight}{Undefined option to highlight: #1}{}]%
}%

\newcommand{\strikenOut}[2]{%
    \IfEqCase{#1}{%
        {color}{\st{#2}}%
        {clean}{}%
    }[\PackageError{strikenOut}{Undefined option to strikenOut: #1}{}]%
}%

\begin{document}

\title{Optimizing Edge Offloading Decisions \\ for Object Detection}

\makeatletter
\newcommand{\newlineauthors}{%
  \end{@IEEEauthorhalign}\hfill\mbox{}\par
  \mbox{}\hfill\begin{@IEEEauthorhalign}
}
\makeatother

\author{\IEEEauthorblockN{Jiaming Qiu\IEEEauthorrefmark{1}}
\and
\IEEEauthorblockN{Ruiqi Wang\IEEEauthorrefmark{1}}
\and
\IEEEauthorblockN{Brooks Hu\IEEEauthorrefmark{2}}
\and
\IEEEauthorblockN{Roch Gu\'{e}rin\IEEEauthorrefmark{1}}
\and
\IEEEauthorblockN{Chenyang Lu\IEEEauthorrefmark{1}}
\newlineauthors
\IEEEauthorblockA{\IEEEauthorrefmark{1}\textit{Washington University in St. Louis}\\
St. Louis, USA\\
\{qiujiaming, ruiqi.w, guerin, lu\}@wustl.edu}
\and
\IEEEauthorblockA{\IEEEauthorrefmark{2}\textit{Northwestern University}\\
Evanston, USA\\
brookshu2026@u.northwestern.edu}
\thanks{This work was supported in part by NSF under grant 2006530 (CNS) and by the Fullgraf Foundation.}}

\maketitle
\thispagestyle{plain}
\pagestyle{plain}

\begin{abstract}
Recent advances in machine learning and hardware have produced embedded devices capable of performing real-time object detection with commendable accuracy. We consider a scenario in which embedded devices rely on an onboard object detector, but have the option to offload detection to a more powerful edge server when local accuracy is deemed too low. Resource constraints, however, limit the number of images that can be offloaded to the edge.  Our goal is to identify which images to offload to maximize overall detection accuracy under those constraints.  To that end, the paper introduces a reward metric designed to quantify potential accuracy improvements from offloading individual images, and proposes an efficient approach to make offloading decisions by estimating this reward based only on local detection results.  The approach is computationally frugal enough to run on embedded devices, and empirical findings indicate that it outperforms existing alternatives in improving detection accuracy even when the fraction of offloaded images is small. Code for the paper's solution is available at \url{https://github.com/qiujiaming315/edgeml-object-detection}.
\end{abstract}


\section{Introduction}
\label{sec:introduction}

The past decade witnessed rapid advances in computer vision and its application to object detection. Driven by innovative algorithm designs and architectural enhancements, contemporary object detection methodologies~\cite{NIPS2015_14bfa6bb, He_2017_ICCV, Lin_2017_CVPR} harnessed deep learning to achieve remarkable improvements in detection accuracy. Concurrently, many object detectors were also designed with inference speed in mind, catering to the exigencies of real-time applications~\cite{yolov3, ssd, Tan_2020_CVPR}. Those advances are increasingly available to and deployed in embedded systems.

Resource constraints in embedded systems, however, commonly call for object detection models with a lighter computational footprint, \ie \emph{weak} models.  This can result in an occasional reduction in accuracy. Edge computing offers a powerful solution to mitigate such reductions by leveraging resources of nearby edge servers accessible over local networks.  Specifically, edge servers can run \emph{strong} models with superior detection accuracy, to which embedded devices offload images when local detection results are deemed insufficiently accurate. 
Local-edge collaborations that rely on such weak-strong combinations have been successfully used in a range of
applications. Recent examples include 
anomaly detection and alert actuation in intelligent Internet of Things (IoT)~\cite{kim2021deep}, and real-time video analytics~\cite{nalaie2022deepscale, ghosh2023react, tan2021deep, hanyao2021edge, deng2020fedvision} in tasks as diverse as surveillance, traffic monitoring, and industrial automation.

A problem of interest is then to, given a certain ``offloading budget'', find the images whose offloading 
maximizes overall detection accuracy. Answering this question is non-trivial.  This is because the accuracy of object detectors depends on multiple factors, \eg whether the model accurately identifies the number and locations of all objects, whether each detected object is correctly labeled with the right class, etc.  This makes assigning a precise measure of the accuracy gains from offloading a given image complex. Matters are further complicated by the fact that, when making offloading decisions, embedded devices have access to neither ground truth annotations nor the strong detector's prediction results. This makes it difficult to devise a reliable yet practical
metric for assessing when offloading an image to the strong detector may improve accuracy.

\begin{table}[t]
\begin{center}
\caption{Comparison with earlier works}
\label{tab:comparison}
\begin{tabular}{|c|c|c|c|}
\hline
& \makecell{\textbf{defined} \\ \textbf{through mAP}} & \makecell{\textbf{evaluated} \\ \textbf{with context}} & \makecell{\textbf{adaptive to} \\ \textbf{offloading budget}} \\
\hline
\makecell{\textbf{Adaptive} \\ \textbf{Feeding~\cite{zhou2017adaptive}}} & \cmark & \xmark & \xmark \\
\hline
\textbf{DCSB~\cite{cao2023edge}} & \xmark & \xmark & \xmark \\
\hline
\makecell{\textbf{ORIC} \\ 
\textbf{(this paper)}} & \cmark & \cmark & \cmark \\
\hline
\end{tabular}
\end{center}
\vspace{-0.5cm}
\end{table}

The most relevant proposals are those of~\cite{zhou2017adaptive, cao2023edge}, whose metrics map to 
thresholds for making offloading decisions. The relevance of those approaches notwithstanding, they suffer from two shortcomings. First, while both seek to evaluate the benefits of offloading individual images, neither account for the broader \emph{context} of that decision.  We expand on this notion in Section~\ref{sec:background}, but it reflects the fact that offloading an image can affect detection accuracy across multiple object classes, including classes not present in the image. Neither of the metrics of~\cite{zhou2017adaptive, cao2023edge} account for that aspect. The second shortcoming of both approaches originates from designs that impose fixed threshold values. This results in policies that are unable to adjust offloading ratios at runtime in response to changes in requirements or operating conditions.

\begin{figure*}[!t]
\centering
\begin{subfigure}{0.2394\linewidth}
  \centering
  \includegraphics[width=\linewidth]{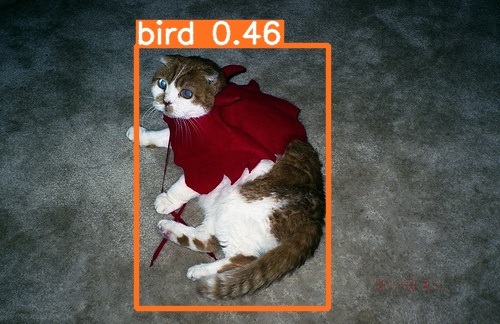}
  \caption{classification error}
  \label{fig:mis_clasification}
\end{subfigure}
\begin{subfigure}{0.2325\linewidth}
  \centering
  \includegraphics[width=\linewidth]{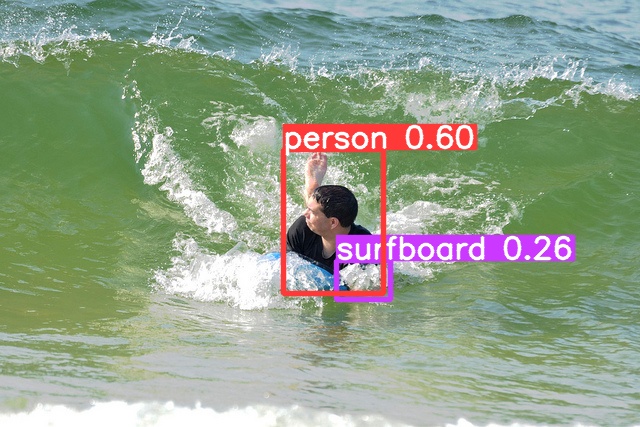}
  \caption{localization error}
  \label{fig:mis_localization}
\end{subfigure}
\begin{subfigure}{0.2068\linewidth}
  \centering
  \includegraphics[width=\linewidth]{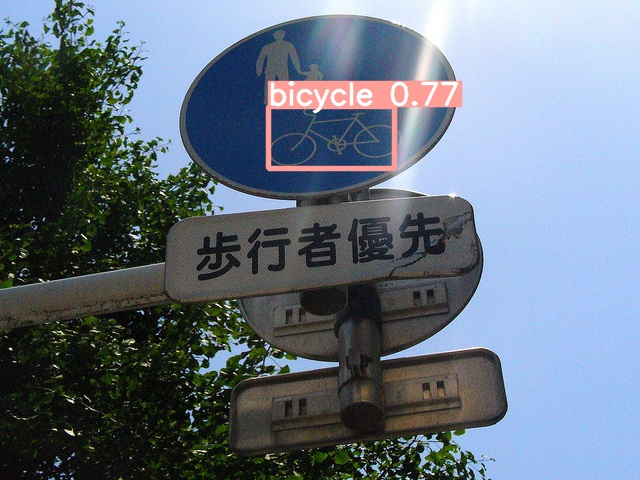}
  \caption{background error}
  \label{fig:mis_background}
\end{subfigure}
\begin{subfigure}{0.2713\linewidth}
  \centering
  \includegraphics[width=\linewidth]{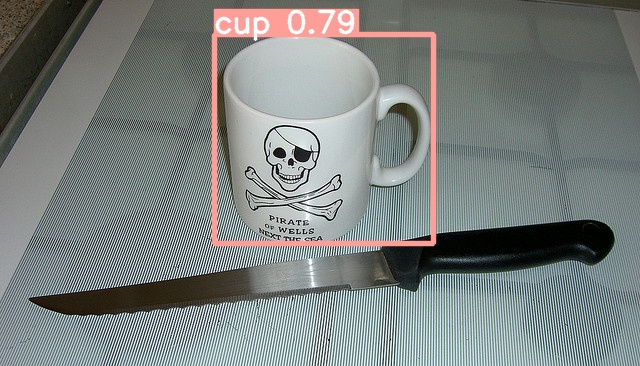}
  \caption{miss ground truth error}
  \label{fig:mis_detection}
\end{subfigure}
\caption{Object detection results that suffer from: (a) Classification error, where a cat is mis-classified as a bird. (b) Localization error, where the predicted location of the surfboard deviates from the ground truth. (c) Background error, where a hallucinated bicycle is detected. (d) Missed ground truth error, where the knife is not detected.}
\label{fig:error_example}
\end{figure*}

Towards addressing these issues, the paper introduces an offloading reward metric, Object detection Reward per Image within
Context (ORIC). ORIC quantifies, for a given image, the difference in 
detection accuracy between weak and strong detectors
by evaluating its impact on a \emph{representative} image subset, \ie the context.  This context captures how offloading an image affects detection accuracy not only for object classes identified in that image, but also in all classes present in a representative subset, \highlight{\useColor}{sampled from the training set of the object detector. Note that, our approach does not require extra data since the context is entirely extracted from the object detector training set.} ORIC aligns 
with the canonical object detection accuracy metric, mean Average Precision (mAP)~\cite{padilla20,medium2018map}, and offers an accurate prediction
of the anticipated mAP enhancement from offloading an individual image. Just as important as the definition of the metric itself, we propose an efficient approach for embedded devices to predict ORIC using only the output of the weak detector. \highlight{\useColor}{The approach is applicable to common local-edge collaboration paradigms based on a cascade of models, such as early exiting~\cite{teerapittayanon2016branchynet, teerapittayanon2017distributed, laskaridis2020spinn, zeng2019boomerang}.}

Because ORIC captures the actual improvement in mAP associated with individual offloading decisions, those decisions can easily be adapted at runtime to any offloading budget.
For example, if limits in network bandwidth or edge server capacity call for no more than $20\%$ of images to be offloaded, this can be readily realized when using ORIC by only offloading images whose ORIC-based predicted reward is in the top $20\%$.

Table~\ref{tab:comparison} compares ORIC to the approaches of~\cite{zhou2017adaptive, cao2023edge} along the dimensions of (i) consistency with mAP, (ii) accounting for the impact of offloading decisions in a broader context than a single image, and (iii) the ability to adjust offloading ratios at runtime, with Section~\ref{sec:background} presenting a more detailed discussion.

In summary, the paper's contributions include:

\begin{itemize}[leftmargin=*,nosep]
    \item A reward metric, ORIC, that accurately identifies which images to offload
    to ``best'' improve mAP, the canonical metric of detection accuracy.
    \item A practical and efficient machine learning model to estimate ORIC based only on information from the weak detector.    
    \item A comprehensive evaluation spanning several object detector pairs and image datasets, and benchmarking on a realistic edge computing testbed\footnote{We rely on publicly available datasets, Microsoft COCO (COCO)~\cite{lin2014microsoft} and Pascal VOC (VOC)~\cite{everingham2010pascal}, and the code required to reproduce the results is available at \url{https://github.com/qiujiaming315/edgeml-object-detection}.
}.
\end{itemize}

The rest of the paper is structured as follows: Section~\ref{sec:background} provides a brief background on object detection and introduces our research motivations. Section~\ref{sec:formulation} formally formulates the problem we tackle, with Section~\ref{sec:definition} expounding on our offloading reward metric ORIC.
Section~\ref{sec:estimation} presents our approach to estimating ORIC, 
while its evaluation is carried out in Section~\ref{sec:evaluation} for different combinations of weak and strong detectors.
Finally, Section~\ref{sec:related_work} discusses related work, and Section~\ref{sec:conclusion} summarizes the paper's findings and contributions.

\section{Background}
\label{sec:background}

Object detection is a classic computer vision technique that aims at recognizing semantic objects within images. It is a crucial technology for diverse applications from autonomous vehicles identifying obstacles, to retail systems monitoring stock on shelves. It differs from image classification, whose goal is to categorize an image's overall content, with object detection aiming to capture all instances of objects that appear in an image. This distinction adds layers of complexity, necessitating models that not only classify objects but also localize them within the image. Localization is commonly realized by proposing bounding boxes of rectangular shapes that encompass the detected objects.  When evaluating the accuracy of object detection, it is, therefore, vital to consider aspects beyond mere classification accuracy. Specifically, in addition to classification error, potential pitfalls for object detectors include misidentifying an object's position (\ie localization error), erroneously detecting background elements as objects (\ie background error), or overlooking certain objects (\ie missed ground truth error). \fig{fig:error_example} shows examples of those typical types of errors in object detection.

\begin{figure*}[!t]
\centering
\begin{subfigure}{0.1710\linewidth}
  \centering
  \includegraphics[width=\linewidth]{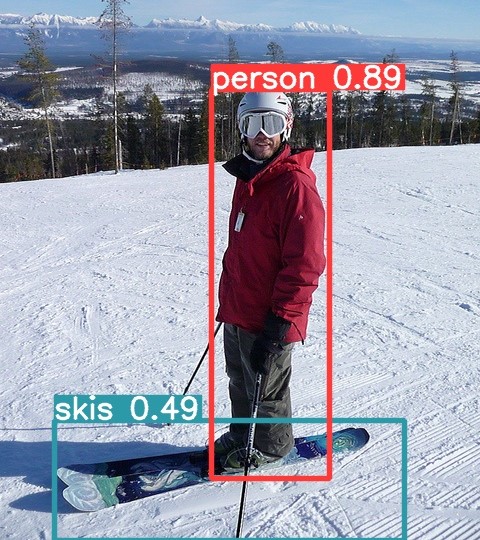}\\
  \vspace{0.1cm}
  \includegraphics[width=\textwidth]{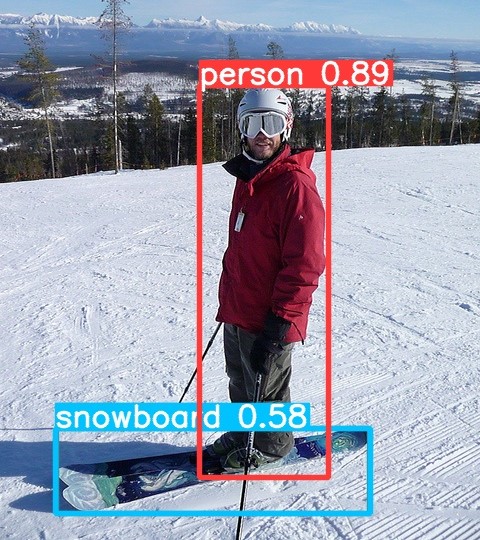}
  \caption{weak is better}
  \label{fig:reward_negative}
\end{subfigure}
\begin{subfigure}{0.2404\linewidth}
  \centering
  \includegraphics[width=\linewidth]{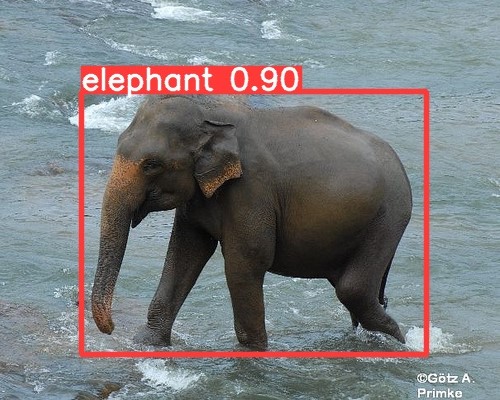}\\
  \vspace{0.1cm}
  \includegraphics[width=\textwidth]{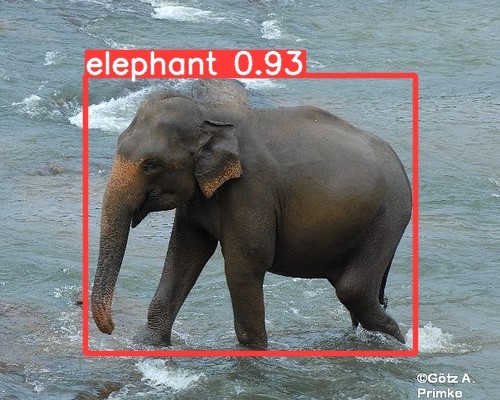}
  \caption{two are equally good}
  \label{fig:reward_zero}
\end{subfigure}
\begin{subfigure}{0.2884\linewidth}
  \centering
  \includegraphics[width=\linewidth]{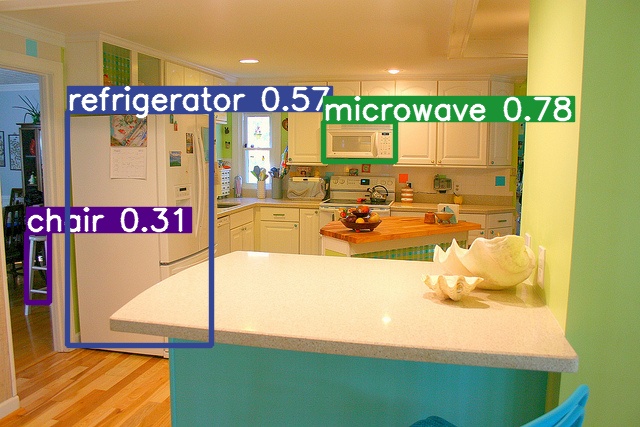}\\
  \vspace{0.1cm}
  \includegraphics[width=\textwidth]{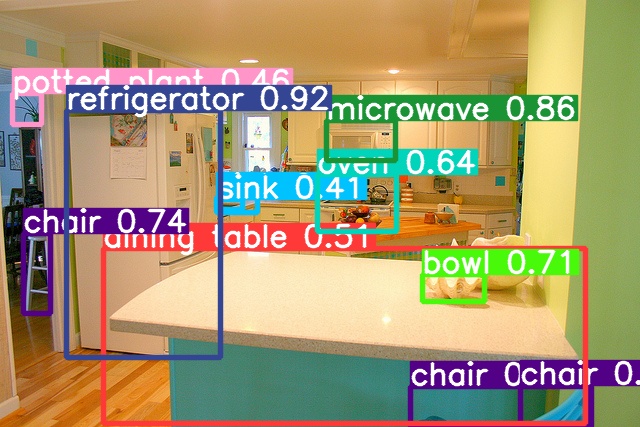}
  \caption{strong is better}
  \label{fig:reward_positive}
\end{subfigure}
    \caption{Images on which the weak detector’s (YOLOv5n) prediction (Top) is considered to be (a) better than, (b) the same with, or (c) worse than the strong detector’s (YOLOv5m) prediction (Bottom).}
    \label{fig:reward_example}
\end{figure*}

In an edge computing setting where local devices equipped with a weak detector can offload images to an edge server with a strong(er) detector, judicious selection of which images to offload can crucially affect the system's overall detection accuracy. 
This is because the performance of the two detectors can vary greatly across images.  \fig{fig:reward_example} showcases this diversity using detection results from YOLOv5n (weak model) and YOLOv5m (strong model) on three images sourced from Microsoft COCO~\cite{lin2014microsoft}. In\highlight{\useColor}{~\fig{fig:reward_negative}}, the weak detector outperforms the strong detector that makes a classification error. In\highlight{\useColor}{~\fig{fig:reward_zero}} both detectors give similar predictions, while in\highlight{\useColor}{~\fig{fig:reward_positive}} the strong detector identifies numerous objects overlooked by the weak detector. Offloading either of the first two images would be a waste and could even degrade detection accuracy. In contrast, offloading the third image improves detection accuracy. Judicious offloading decisions, therefore, call for a carefully calibrated metric aimed at 
quantifying the accuracy enhancement potential from offloading individual images.
Mean Average Precision (mAP) is a canonical metric for object detection accuracy, with widespread acceptance within the computer vision domain~\cite{padilla20,medium2018map}. It proceeds by first collecting all bounding boxes proposed by the object detector for each predicted class, before computing a precision-recall curve based on those bounding boxes sorted by their confidence scores.  The results are then summarized using the Average Precision (AP) metric. The final mAP is derived by averaging the AP over all $M$ object classes.  mAP is normally used to evaluate the accuracy of an object detector on an \emph{entire} dataset. Nonetheless, \cite{zhou2017adaptive} introduced a metric termed mean Average Precision \emph{per Image} (mAPI), which evaluates the mAP of an object detector with respect to a \emph{single} image $i$:
\beq
\label{eqt:mapi}
mAPI_i = \frac{1}{S} \sum_{j=1}^S AP_{ij},
\eeq
where $AP_{ij}$ is the AP of class $j$ in image $i$ and $S$ is the number of classes within image $i$, usually a subset of the $M$ classes of the full dataset.

Using mAPI, \cite{zhou2017adaptive} introduced a simple offloading reward metric based on the difference in mAPI values between the strong and weak detectors for the image under consideration. We term this metric Object detection Reward per Image (ORI) and use it as a baseline against which to compare ORIC. 

\begin{figure}[t]
    \centering
\begin{subfigure}{0.49\linewidth}
  \centering
  \includegraphics[width=0.7\linewidth]{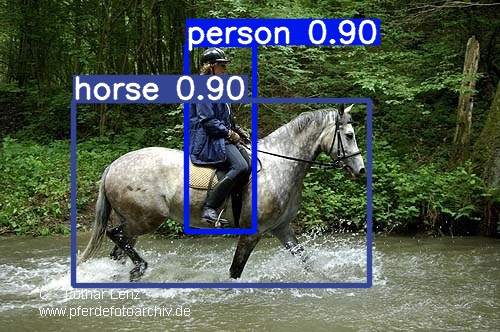}\\
  \vspace{0.1cm}
  \includegraphics[width=\linewidth]{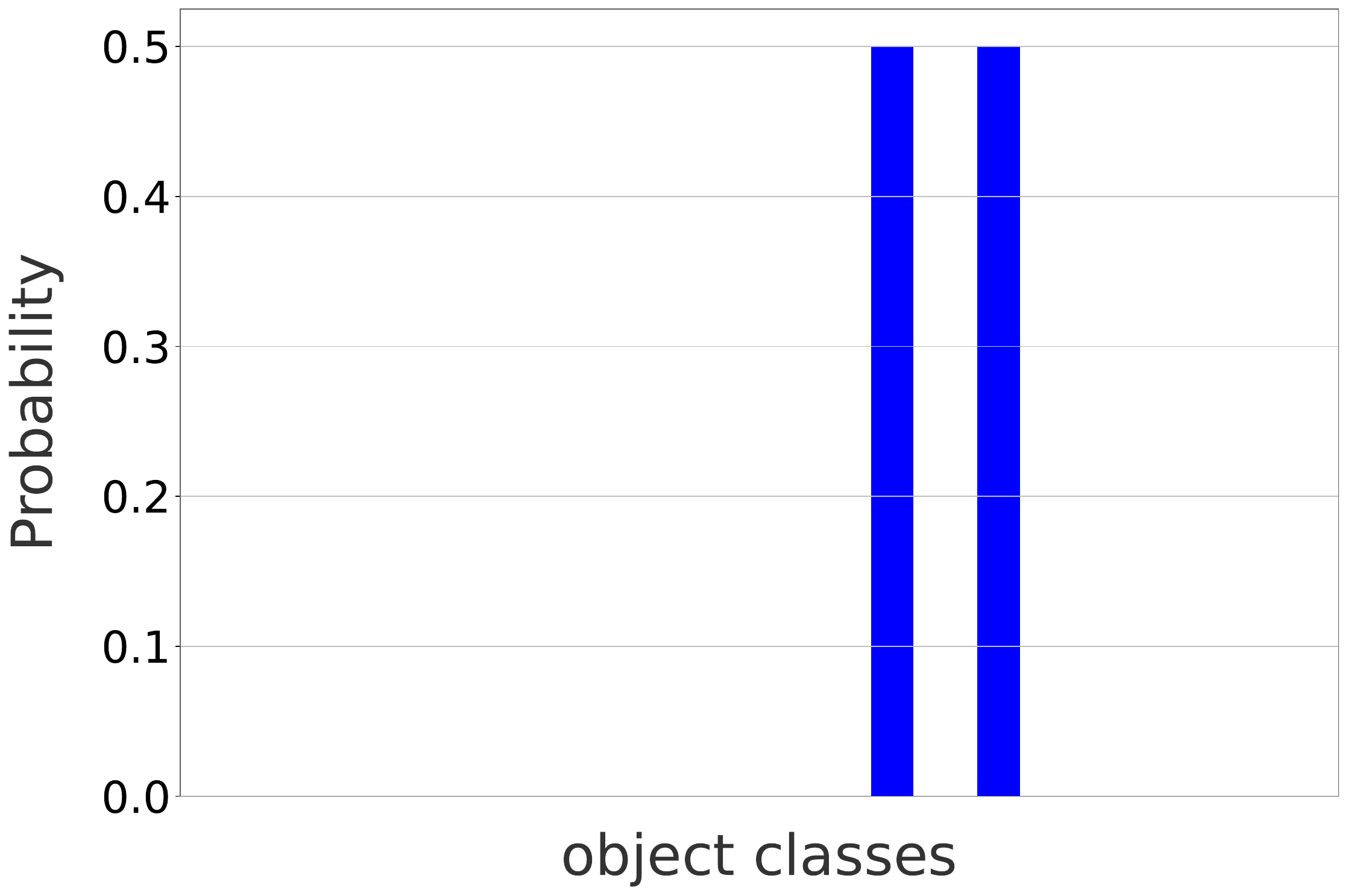}
  \caption{sample image}
  \label{fig:sample_image}
\end{subfigure}
\begin{subfigure}{0.49\linewidth}
  \centering
  \includegraphics[width=0.7\linewidth]{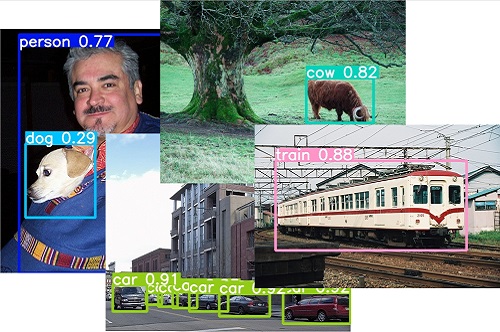}\\
  \vspace{0.1cm}
  \includegraphics[width=\linewidth]{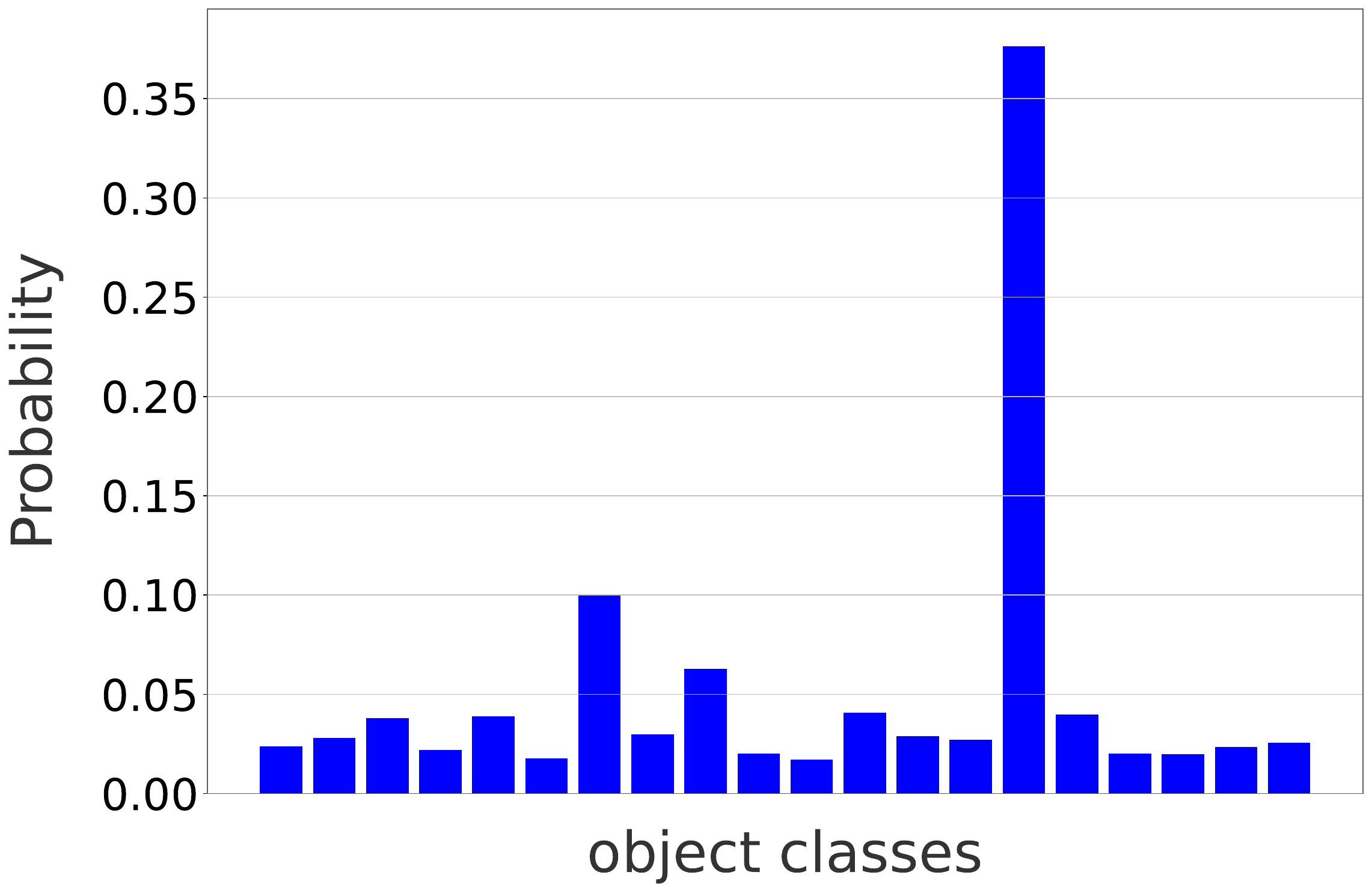}
  \caption{Pascal VOC dataset}
  \label{fig:Pascal_dataset}
\end{subfigure}
\caption{Object classes frequency in (a) representative image, and (b) across the Pascal VOC dataset~\cite{everingham2010pascal}.}
    \label{fig:class_dist}
\end{figure}

Recalling our earlier comment on the importance of \emph{context}, we note that mAPI (ORI) only evaluates AP for true classes in image~$i$ and ignores other classes.  This overlooks certain types of errors. Consider a scenario where the only object in image~$i$ is a cat, yet the weak detector, while accurately predicting the presence of the cat, makes a background error by hallucinating a dog. An mAP evaluation over the entire dataset (that includes images of dogs and cats) would count the dog prediction as detrimental to the AP of the dog class. However, because mAPI only considers image~$i$, it overlooks this error (image~$i$ lacks a dog as a true class, so that both precision and recall of the dog class are $0$). This result in mAPI misjudging the mAP improvement from offloading image~$i$.

\highlight{\useColor}{We expand on the potential impact of such errors in~\fig{fig:class_dist}, which uses the Pascal VOC dataset~\cite{everingham2010pascal} to show the number of (ground truth) object classes in a single, representative image (\fig{fig:sample_image}), and the corresponding distribution across all 20~object classes present in the entire dataset (\fig{fig:Pascal_dataset}).  The figures indicate that individual images tend to have much fewer object classes than the entire dataset.  Hence, as the previous example sought to illustrate, because the $S$ classes of \Eqref{eqt:mapi} only account for classes present in the image under consideration, this can bias mAPI-based reward estimates.}

The approach of DCSB~\cite{cao2023edge} suffers from similar limitations.  It was motivated by a heuristic that favors offloading images with undetected objects, and, therefore, relies on
a simple (binary) offloading reward metric based on whether the strong detector detects more objects than the weak detector.  This is not only a rather coarse metric, but it also again fails to account for possible hallucination errors by the weak detector.

ORIC, by incorporating the effect of offloading an image in the \emph{context} of a representative set of images, seeks to overcome these limitations and to offer a more accurate estimate of the resulting mAP improvements.  We expand on this approach and its benefits in the next sections.

\section{Problem Formulation}
\label{sec:formulation}

\begin{figure*}[!t]
\centering
    \includegraphics[width=0.7\textwidth]{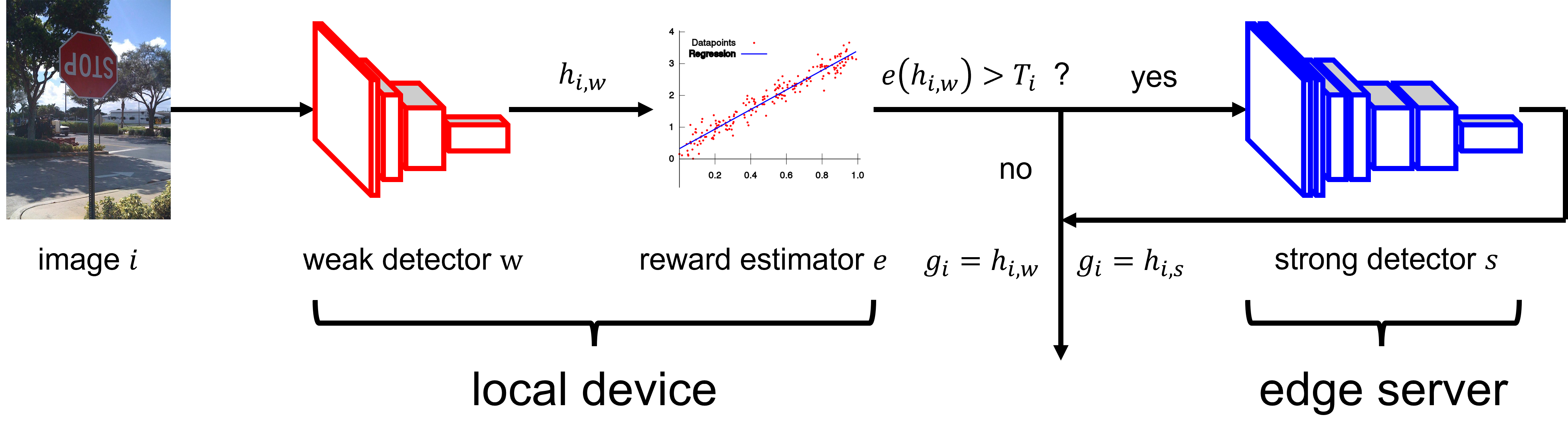}
    \caption{Image offloading pipeline. Each image $i$ is first processed by the weak detector $w$ on the local device, based on whose detection results $h_{i,w}$ the reward estimator makes a prediction $e(h_{i, w})$ on the offloading reward. The system decides whether to offload the image to the strong detector $s$ on the edge by comparing the reward estimate with a threshold $T_i$.}
    \label{fig:system_diagram}
\end{figure*}

Consider an edge computing setting where local devices collect and perform object detection.
Local devices can offload selected images to an edge server for better detection results. Due to more limited computational resources, the local object detector is denoted as the \emph{weak} detector versus the \emph{strong} detector available on the edge server. We assume that, across a given set of $N$ images, the strong detector is more accurate and achieves a higher mAP on that set:
\begin{equation*}
\map(\bigcup_{1 \leq i \leq N}h_{i, w}) < \map(\bigcup_{1 \leq i \leq N}h_{i, s})\, ,
\end{equation*}
where, for each image $i$, $h_{i, w}$ and $h_{i, s}$, are the detection results \highlight{\useColor}{(\ie the set of predicted bounding boxes)} of the weak and strong detectors, respectively.
The two are trained over a dataset that shares the same set of object classes and input image distribution as the dataset used for evaluation at runtime.

Our goal is to find a subset of images $\mathcal{S}^*$ that achieves the maximum mAP enhancement when offloaded:
\beq
\begin{aligned}
\label{eq:s*}
\mathcal{S}^* &= \argmax_{\mathcal{S} \in \mathcal{C}}\big\{\map(\bigcup_{1 \leq i \leq N}g_{i})\big\},\\
g_i &= 
\begin{cases}
    h_{i, w}, &i \not\in \mathcal{S}\\
    h_{i, s}, &i \in \mathcal{S}\\
\end{cases},
\end{aligned}
\eeq
where $\mathcal{C}$ encompasses all subsets $\mathcal{S}$ that satisfy certain resource constraints. For example, if at most 10\% images can be offloaded to avoid overloading the network or the edge server, then $\mathcal{C} = \big\{\mathcal{S}: |\mathcal{S}|/N \leq 0.1 \big\}$.

Exploring all candidate $\mathcal{S} \in \mathcal{C}$ is combinatorial in nature, and quickly becomes impractical as $N$ scales. Moreover, in practical scenarios, local devices typically process images on the fly, which means neither results on the complete image set, nor the relative gains afforded by the strong detector are accessible when making image offloading decisions. 

In light of these constraints, we introduce an offloading reward metric $R_i$ that seeks to measure the incremental improvement in mAP when image $i$ is offloaded ($i \in \mathcal{S}$) compared to not offloaded ($i \not\in \mathcal{S}$). We introduce this reward metric more formally in Section~\ref{sec:definition}, but its primary purpose is to convert the complex combinatorial problem of~\Eqref{eq:s*} into a more tractable one that supports individual offloading decisions.  Computing this reward metric, however, involves a comparison of two mAP values. One when the image is offloaded and one when it is not, with the former requiring the result of the strong detector. This result is obviously not available when an offloading decision needs to be made.  As a result, it is also necessary to define a reward estimator $e(h_{i, w})$ to predict the offloading reward $R_i$ using only the weak detector's result $h_{i, w}$.  The resulting setup is illustrated in\marginpar[]{}\highlight{\useColor}{~\fig{fig:system_diagram}}, where the local device offloads image~$i$ only if the reward estimate $e(h_{i, w})$ exceeds a designated threshold $T_i$. 

\highlight{\useColor}{As in~\cite{chakrabarti2021real}, when image~$i$ arrives, the threshold $T_i$ used by the local device can be adjusted to, say, enforce dynamic resource constraints.
} However, for simplicity and clarity of exposition, we assume \highlight{\useColor}{static resource constraints, and therefore a} fixed threshold policy, \ie $T_i = T$. For a given offloading ratio $r$, $T$ is then the $(1 - r)^{th}$ percentile of the reward estimates, and $\mathcal{S}$ is the subset of images with reward estimates above $T$.

Optimizing the system's mAP, therefore, calls for finding a suitable pair of image offloading reward metric, $R^*$,  and estimator, $e^*(\cdot)$. This yields the following formulation:
\beq
\label{eq:r*}
\begin{aligned}
(R^*, e^*) &= \argmax_{R, e}\big\{\map(\bigcup_{1 \leq i \leq N}g_{i})\big\},\\
g_i &= 
\begin{cases}
    h_{i, w}, &e(h_{i,w}) \leq T\\
    h_{i, s}, &e(h_{i,w}) > T\\
\end{cases}.
\end{aligned}
\eeq

Developing a solution to \eqref{eq:r*} calls for addressing several underlying challenges:
\begin{itemize}[leftmargin=*,nosep]
    \item mAP as a metric aggregates detection results spanning an entire image set. mAP's improvement from offloading a single image is, therefore, dependent on other images and their own offloading decisions. Accounting for both is infeasible in practice, and the notion of a (per image) offloading reward is, therefore, likely an approximation.
    \item Even if an accurate per-image reward can be computed, it involves access to $h_{i, w}$, $h_{i, s}$ and the ground truth.  Only $h_{i, w}$ is available when making offloading decisions, and estimating the resulting reward, therefore, implies inferring errors made by the weak detector and not the strong detector.
    \item The quality of a reward-estimator pair, $(R,e)$, depends on how well $R$ captures the effect of individual offloading decisions on mAP, and how feasible it is to devise an efficient estimator $e$ for $R$; a better $R$ may be harder to estimate. The interplay between the two remains elusive.
\end{itemize}
To address those challenges we decouple defining (Section~\ref{sec:definition}) and estimating (Section~\ref{sec:estimation}) $R$. This is arguably a limitation, but our choice of $R$ closely emulates how offloading decisions affect mAP, while our estimator $e$ results from a comprehensive evaluation of different estimation methods to predict~$R$.  

\section{ORIC: Reward Metric for Edge Object Detection}
\label{sec:definition}

Next, we formally define our offloading reward metric, Object detection Reward per Image with Context evaluation (ORIC), and compare it with ORI (Section~\ref{sec:background}) to demonstrate its benefits in making offloading decisions that improve mAP.

\subsection{Definition}
As ORI, 
ORIC seeks to capture differences in mAP produced by individual images. A key distinction is that while ORI involves summing average precision measures across classes for the image under consideration (see \Eqref{eqt:mapi}), ORIC is computed over a \emph{representative} image set $\mathcal{E}$ (the \emph{context}) rather than a \emph{single} image. The primary motivation is to more closely reflect the ensemble average nature of mAP.  To this end, we first introduce an augmented definition of mAPI, termed mAPC (mAP per image through Context evaluation):
\beq
\label{eqt:mapc}
mAPC_i = \map(\{h_i, H_{\mathcal{E}}\}) = \frac{1}{E} \sum_{j=1}^E AP_{\{i\} \cup \mathcal{E}, j}\, ,
\eeq
where $H_{\mathcal{E}} = \bigcup_{1 \leq k \leq |\mathcal{E}|}h_{k}$ denotes the consolidated detection results for the image set $\mathcal{E}$, $E$ is the number of classes across image $i$ and images in $\mathcal{E}$, and $AP_{\{i\} \cup \mathcal{E}, j}$ refers to the AP of class $j$ evaluated on 
$\{h_i\}\cup H_{\mathcal{E}}$.

In other words, given an image~$i$, 
mAPC evaluates the mAP of an object detector on $\{i\}\cup \mathcal{E}$, \ie using not just image~$i$, but also the broader context $\mathcal{E}$.
The offloading reward metric $ORIC_i$ for image~$i$ is then defined as:
\beq
\label{eq:oric}
\begin{aligned}
ORIC_i =& (|\mathcal{E}| + 1)\cdot(mAPC_{i, s} - mAPC_{i, w})\\
=& (|\mathcal{E}| + 1)\cdot\big[\map(\{h_{i, s}, H_{\mathcal{E}, w}\})\\
& - \map(\{h_{i, w}, H_{\mathcal{E}, w}\})\big],
\end{aligned}
\eeq
where $w$ and $s$ indicate the type of detector (weak or strong).

\Eqref{eq:oric} is the difference in mAP between offloading and not offloading image $i$, when evaluating the mAP on $\{i\}\cup \mathcal{E}$. The term $|\mathcal{E}| + 1$ is a scaling factor to normalize ORIC across representative image sets of different sizes. Note that, mAPC (and therefore ORIC) defaults to mAPI (ORI) when $\mathcal{E} = \varnothing$. 

In evaluating the impact of offloading image~$i$ to the strong detector, \Eqref{eq:oric} also assumes that the detector used for images in $\mathcal{E}$ is the weak detector.  This is reasonable since, in most scenarios, the number of images that can be offloaded is small, so that most images are handled by the weak detector.  We experimented with using the strong detector or a mixture of the two, but there was limited sensitivity to this choice.  

The choice of which images to include in $\mathcal{E}$ is of greater significance.  
\highlight{\useColor}{We need to select $\mathcal{E}$ to ensure it is representative of the entire dataset, while keeping computational costs, and hence the cardinality of $\mathcal{E},$ low\footnote{\highlight{\useColor}{As seen in \Eqref{eqt:mapc}, this cost depends on $AP_{\{i\}\cup\mathcal{E},j}$, which grows with $|\mathcal{E}|$, the cardinality of $\mathcal{E}$.}}.  To that end,}
we curate $\mathcal{E}$ by uniformly sampling without replacement images from the complete dataset. \highlight{\useColor}{The goal is to collect an unbiased estimate of the object class distribution across the dataset (see~\fig{fig:Pascal_dataset}) using a relatively small sample.}

\highlight{\useColor}{As alluded to, the motivation for ORIC's context $\mathcal{E}$, is to better estimate how detection results for an individual image affect the averaging that underlies mAP's computation.  The role of $\mathcal{E}$ is to reflect the relative weights of object classes when evaluating the impact of the ``competing'' detection decisions of the weak and strong detectors for that image.}

In the rest of this section, we compare the performance of ORIC and ORI assuming access to an offloading reward oracle. That is, $h_{i, w}$, $h_{i, s}$, and the ground truth are available, so that we can compute exact reward values for ORI and ORIC when making offloading decisions for each image.

\subsection{ORIC as an Offloading Metric}
\label{sec:oric_eval}
\begin{figure}[t]
    \centering
    \includegraphics[width=\linewidth]{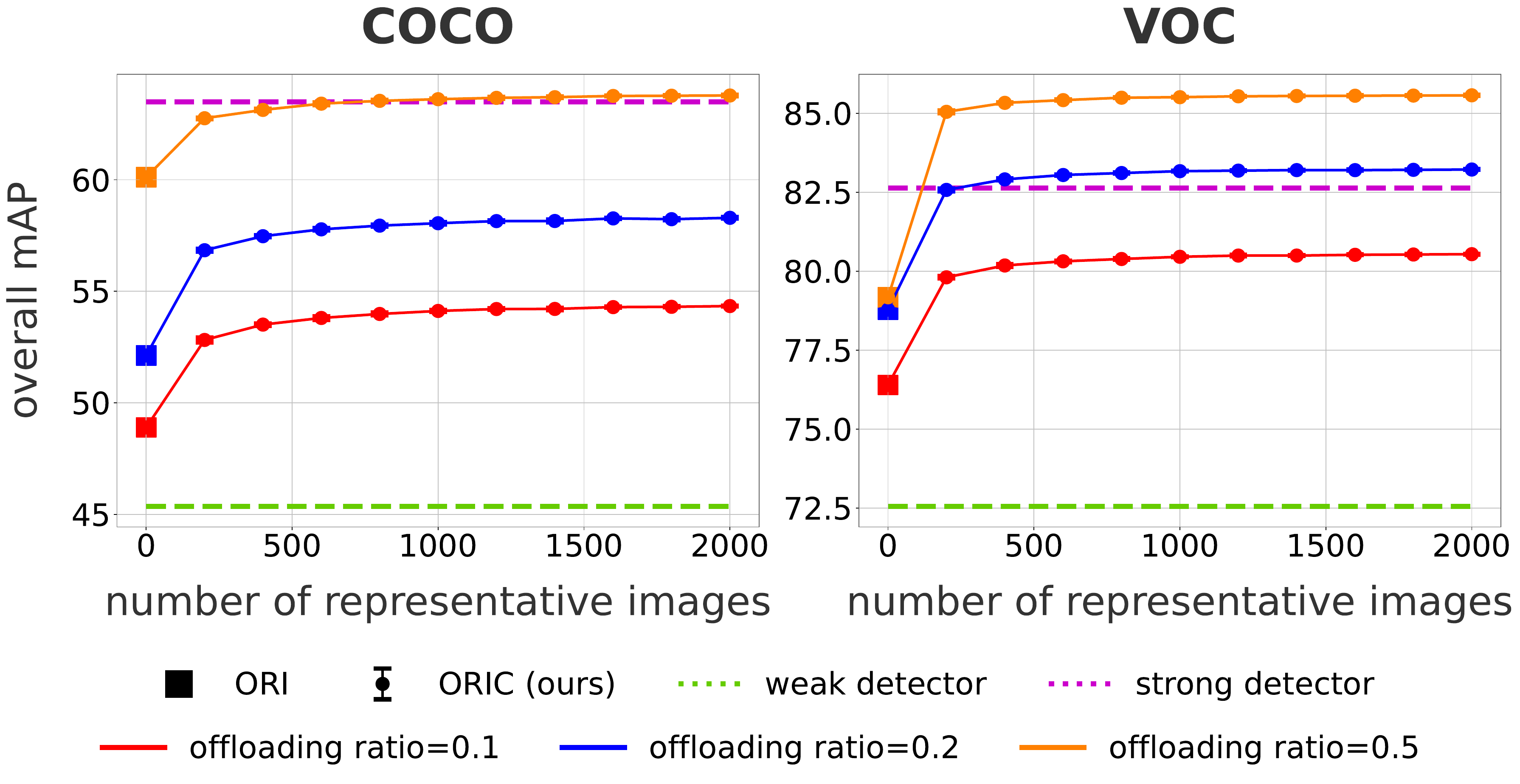}
    \caption{Overall mAP with ORIC and ORI as the offloading reward metrics. ORI assumes $|\mathcal{E}|=0$, while ORIC's performance varies as a function of the context size $|\mathcal{E}|$.}
    \label{fig:map_representation_size}
\end{figure}

\begin{table*}[t]
\renewcommand{\arraystretch}{1.3}
\begin{center}
\caption{Weak and strong detector's mAP on images with positive and non-positive offloading rewards}
\label{tab:image_class_map}
\begin{tabular}{|c|c|ccc|ccc|}
\hline
& & \multicolumn{3}{c|}{\textbf{ORIC (ours)}} & \multicolumn{3}{c|}{\textbf{ORI}} \\
\hline
\multirow{2}{*}{\textbf{Dataset}} & \multirow{2}{*}{\textbf{Image Category}} & \multirow{2}{*}{\textbf{Percentage}} & \multicolumn{2}{c|}{\textbf{mAP}} & \multirow{2}{*}{\textbf{Percentage}} & \multicolumn{2}{c|}{\textbf{mAP}}\\
& & & \textbf{weak} & \textbf{strong} & & \textbf{weak} & \textbf{strong}\\
\hline
\multirow{2}{*}{COCO} & reward $\leq 0$ & $19.44\%$ & $49.98$ & $48.37$ & $42.06\%$ & $61.72$ & $71.46$\\
& reward $ > 0$ & $80.56\%$ & $44.84$ & $67.68$ & $57.94\%$ & $38.52$ & $60.17$\\
\hline
\multirow{2}{*}{VOC} & reward $\leq 0$ & $27.67\%$ & $70.38$ & $63.28$ & $76.76\%$ & $82.07$ & $86.04$\\
& reward $ > 0$ & $72.33\%$ & $73.81$ & $89.77$ & $23.24\%$ & $52.38$ & $74.81$\\
\hline
\end{tabular}
\end{center}
\end{table*}
In exploring ORIC as a metric for offloading decisions, we focus on two aspects.  The first is how the choice of $\mathcal{E}$ influences accuracy.
As the addition of $\mathcal{E}$ is the major distinction between ORIC and ORI, 
understanding how large an $\mathcal{E}$ is needed to claim most of its benefits 
is our initial focus.
\highlight{\useColor}{According to our discussion on~\Eqref{eqt:mapc}, a small $|\mathcal{E}|$ is desirable to keep the computational cost of ORIC low.}
Also of interest is exploring whether ORIC can successfully capture the relative strengths and weaknesses of the weak and strong detectors across images.  This is our second focus, and we demonstrate how ORIC enables us to get the best of both detectors, and outperforms each of them individually.

\subsubsection{Impact of $\mathcal{E}$}
We first investigate the impact of $|\mathcal{E}|$ on the performance of ORIC. 
\fig{fig:map_representation_size} reports the overall mAP achieved by both ORI and ORIC for different offloading ratios\footnote{The images with the highest $r$\% offloading rewards are offloaded.} $r=0.1,0.2,0.5$, and, for ORIC, how it changes as a function of the context size $|\mathcal{E}|$.  
As previously mentioned, $\mathcal{E}$ is generated by random selection without replacement from the validation sets of the two datasets: Microsoft COCO (COCO)~\cite{lin2014microsoft} and Pascal VOC (VOC)~\cite{everingham2010pascal}, used for this experiment. For each value of $|\mathcal{E}|$, the results are based on $10$ separate instances of randomly selecting $\mathcal{E}$, and the figure reports both average mAP values and their 95\% confidence intervals.

For comparison purpose, the figure also reports the mAP value achieved on those datasets by the weak and strong detectors alone.  For this experiment, those detectors were chosen from the YOLO~\cite{yolov3} version 5 suite~\footnote{\url{https://github.com/ultralytics/yolov5}}, with the nano model (YOLOv5n) and the medium model (YOLOv5m), as the weak and strong detectors, respectively. The detectors were trained on training sets from COCO (118,287 images) and VOC (16,551 images). We used their validation sets (5,000 images and 4,952 images for COCO and VOC, respectively) to compute and assess the efficacy of ORI and ORIC. 

The results illustrate the benefit of incorporating a representative set of images, $\mathcal{E}$, in the definition of ORIC.  In particular, as $|\mathcal{E}|$ increases, ORIC quickly outperforms ORI, often by more than $10\%$.  
More interesting is the fact that, at least for our two image sets, COCO and VOC, a relatively small image set $\mathcal{E}$ is sufficient to achieve the bulk of the benefits.  This together with the narrow confidence intervals we observe indicates that our sampling procedure is able to quickly acquire a set of images that accurately capture the distribution and relative importance (to mAP) of object classes in both image sets.  

As mentioned earlier, keeping $|\mathcal{E}|$ small is useful to limit the \highlight{\useColor}{computational cost, which is part of the} training cost of the procedure we use to estimate ORIC. In our experiments, we settled on a value of $|\mathcal{E}|=1,000$, as it realized an acceptable trade-off between accuracy and computational overhead.

\begin{figure*}[t]
    \centering
    \includegraphics[width=0.8\linewidth]{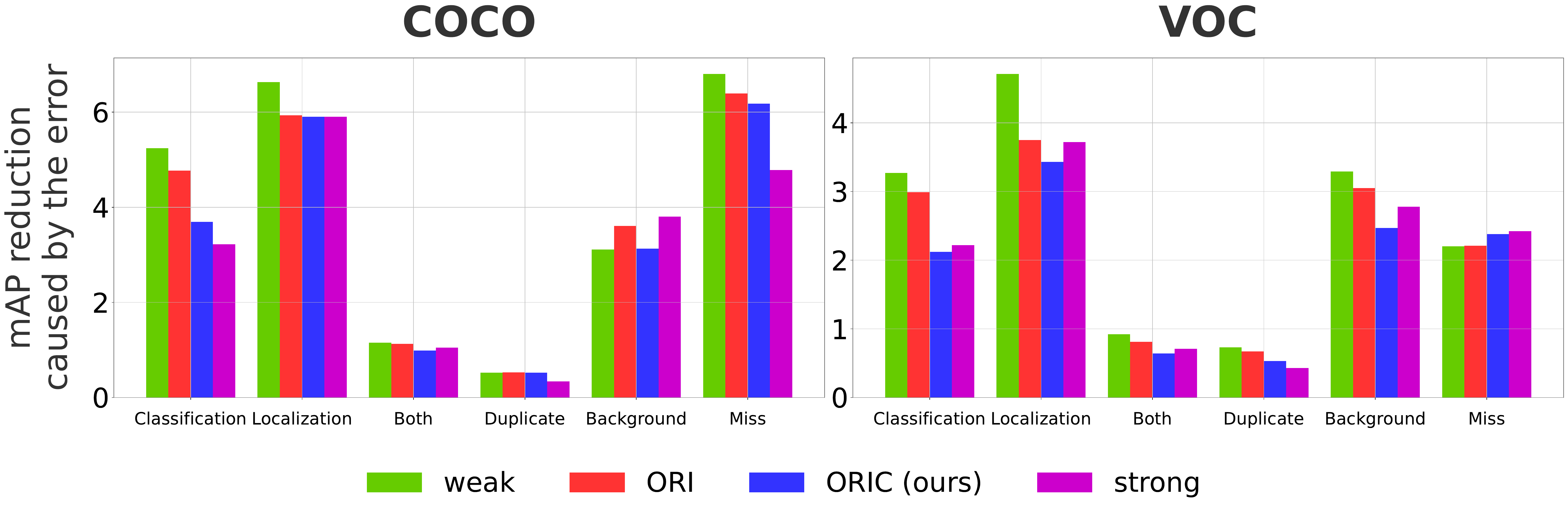}
    \caption{Six types of errors on COCO (Left) and VOC (Right) made by the weak detector, the strong detector, and the edge computing system using oracles of ORIC and ORI as the reward metric with an offloading ratio of $0.2$.}
    \label{fig:error_breakdown}
\end{figure*}

\subsubsection{Outperforming the Strong Detector}
\fig{fig:map_representation_size} hints at another aspect of interest, namely, that an (intelligent) combination of weak and strong detectors can outperform (yield a higher mAP) either.  This is more apparent with the VOC image dataset than with COCO, but holds for both.  This is not unexpected as different detectors often exhibit distinct strengths and weaknesses.

The notion of reward captures those differences, with a positive reward corresponding to images for which the strong detector outperforms the weak detector. Table~\ref{tab:image_class_map} offers further insight into how ORI and ORIC fare in capturing those differences. It categorizes images from the validation sets of COCO and VOC, into positive and non-positive rewards for both ORI and ORIC, and reports the corresponding mAPs for each detector.  The table shows that ORIC is more conservative in its assessment of which images have a non-positive reward, \ie would not benefit from being offloaded ($19.44$\% and $27.67$\% of COCO and VOC images, respectively, in contrast to ORI's $42.06$\% and $76.76$\%). This conservatism is, however, justified as the resulting mAP values indicate.  For those subsets of images, the weak detector outperforms the strong detector.  This is not so when using ORI as the reward metric. 

This illustrates ORIC's superior ability to identify images for which the weak detector outperforms the strong detector.  Although the strong detector still outperforms the weak detector on most images, it is the presence of this subset of images that explains why the ``judicious'' combination of both detectors outperforms either alone.
\highlight{\useColor}{ORIC's ability to better pinpoint images where the weak detector outperforms the strong detector extends to a greater overall accuracy in predicting offloading reward values.  This is the reason for the improvements in mAP (over ORI) observed in \fig{fig:map_representation_size}.}
We investigate more closely the reasons behind ORIC's better performance by comparing, for an offloading ratio of $0.2$, the types of detection errors made by (i) the weak detector, (ii) the strong detector, and (iii)/(iv) combinations of the two when those combinations are driven by reward values computed by ORI (iii) and ORIC (iv) -- images with reward values in the top $20\%$ are assigned to the strong detector, while the remaining ones are handled by the weak detector. As proposed in~\cite{bolya2020tide}, we consider six categories of errors: classification error, localization error, both classification and localization error, duplicate detection error, background error, and missed ground truth error. 

We again use YOLOv5n and YOLOv5m as our detector pair, with~\fig{fig:error_breakdown} reporting measures\footnote{The $y$-axis tracks the amount of mAP reduction, \ie area above the precision-recall curves, caused by each error category.} of the number of errors made by the above four configurations in each of the six categories. From a quick glance at the figure, it is clear that the combination that relies on ORIC for its offloading decisions makes fewer errors across nearly all categories than the one using ORI (the one exception is the ``missed ground truth'' category for VOC, where the difference is minor).  In other words, ORIC's better performance should be reasonably robust across different types of images.

\section{Estimation of ORIC}
\label{sec:estimation}
The previous section evaluated ORIC assuming it could be computed \emph{exactly} with access to the ground truth and results from both the weak and strong detectors.  This will obviously not be the case in practice, where decisions need to be made in real-time based only on the results of the weak detector.  

In this section, we present and evaluate our approach for a practical offloading policy based on ORIC.  It relies on a lightweight \emph{estimator} for ORIC in the form of a simple multi-layer perceptron (MLP) regression model\footnote{The motivation for adopting a regression model rather than a binary classifier as proposed by~\cite{zhou2017adaptive, cao2023edge} is to map our reward estimate to a continuous value space.  This facilitates the use of different offloading thresholds at runtime, \eg to respond to variations in network conditions.} with hyper-parameters fine-tuned via grid search. The model uses only information from the weak detector as its inputs and was trained using the validation sets from COCO (5,000 images) and VOC (4,952 images), with 5-fold cross-validation. It easily runs on embedded systems, and, for each image, the ORIC estimate it generates is passed to a basic threshold-based policy for real-time offloading decisions.

Next, Section\ref{sec:oric_alt} discusses possible \emph{inputs} to the estimator
given that only information from the weak detector is available.  Section~\ref{sec:oric+} builds on Section\ref{sec:oric_alt} and introduces adjustments that improve our ability to accurately estimate ORIC. Finally, Section~\ref{sec:oric_estimate} evaluates the estimator's performance.

\subsection{Input Choices in Estimating ORIC}
\label{sec:oric_alt}

\begin{figure}[t]
    \centering
    \includegraphics[width=0.8\linewidth]{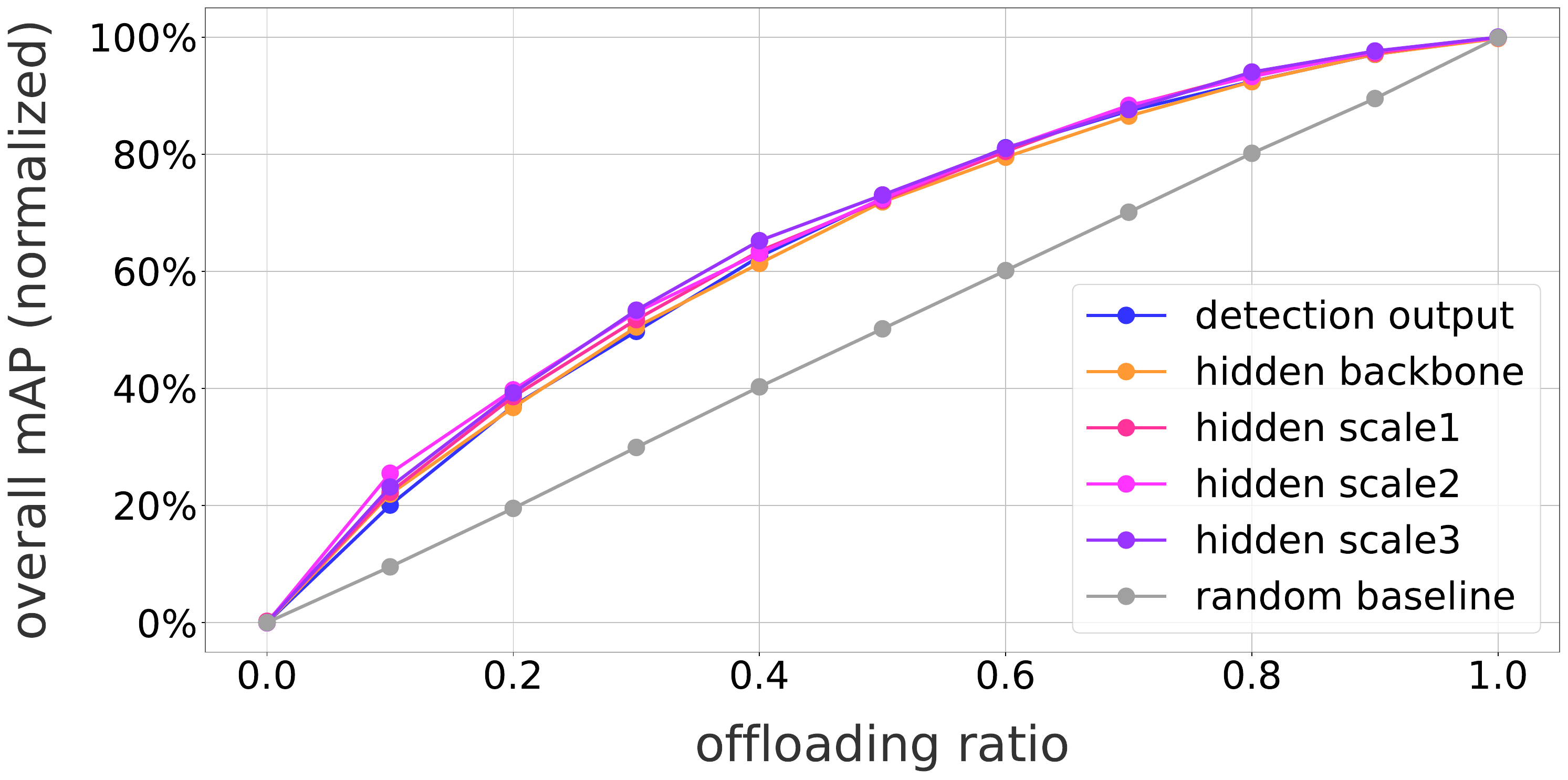}
    \caption{Effect of using different inputs (feature maps from different weak detector's hidden layers and the weak detector's output) to estimate ORIC.}
    \label{fig:input_selection}
\end{figure}

In devising an estimator for ORIC, one that is based only on information available from the weak detector, the first step is to decide what information to use as \emph{inputs} to the estimator.  The output of the weak detector is a natural choice, but richer inputs are possible.  For example, feature maps from various hidden layers of the weak detector may include information that could improve estimation accuracy.  In this section, we explore a few such alternatives.

Our first option uses the output of the weak detector as our input, and, mirroring the methodology presented in~\cite{zhou2017adaptive}, we select features extracted from the weak detector's top~$25$ bounding box proposals, ranked by their confidence scores. 

Next, we evaluated using the following hidden layers of YOLOv5n: the last layer of YOLOv5n's backbone and the last layers of the three detection heads of YOLOv5n. These three detection heads have been tailored to discern objects across varying scales, specifically optimized for the detection of large, medium, and small objects, respectively~\cite{yolov3}. To effectively extract information from the feature maps of these hidden layers, we relied on a lightweight Convolutional Neural Network (CNN) model rather than an MLP as the reward estimator. The hyper-parameter tuning was again executed through grid search for each choice of model input. 

The results are presented in~\fig{fig:input_selection} using COCO 
and YOLOv5n and YOLOv5m as our detector pair. The figure shows that the choice of estimator input has only a limited impact on our ability to accurately estimate ORIC towards improving the system's overall mAP.  Adding the weak detector's hidden layers yields only marginally better performance, and only for low offloading ratios. Given our focus on embedded devices and the need to keep latency and memory footprint low during reward estimation, we elected to rely only on the weak detector's results as inputs to our estimator.

As a side comment, we note that recent edge computing investigations explored partitioning large Deep Neural Network (DNN) models into a head and a tail model for local and edge deployment, respectively. Some of these works investigated attaching early-exits~\cite{teerapittayanon2016branchynet, teerapittayanon2017distributed, laskaridis2020spinn, zeng2019boomerang} to the partitioning point of the DNN model for early prediction outcomes, obviating the need for offloading in instances where local predictions exhibit sufficient confidence. Our exploration reveals that our proposed ORIC-based estimation approach is readily applicable to such emerging edge object detection frameworks with embedded early exits, enabling them to accurately assess the potential gains of offloading even at an early partitioning point, such as the backbone of the object detector.

\subsection{Improving ORIC's Estimation}
\label{sec:oric+}

Recall that ORIC seeks to quantify a reward (improvement in mAP) for offloading individual images.  Any estimator will have limitations, and, in this section, we explore improvements to our basic estimator.  First, given our goal of maximizing mAP's improvement, we decide to emphasize estimation accuracy for high-reward images. Further, and in recognition of the fact that uneven distribution of reward values can make it harder to discriminate between images that should be offloaded and images that should not, we also transform the distribution of ORIC to make such decisions easier. 

\begin{figure}[t]
    \centering
    \includegraphics[width=0.49\linewidth]{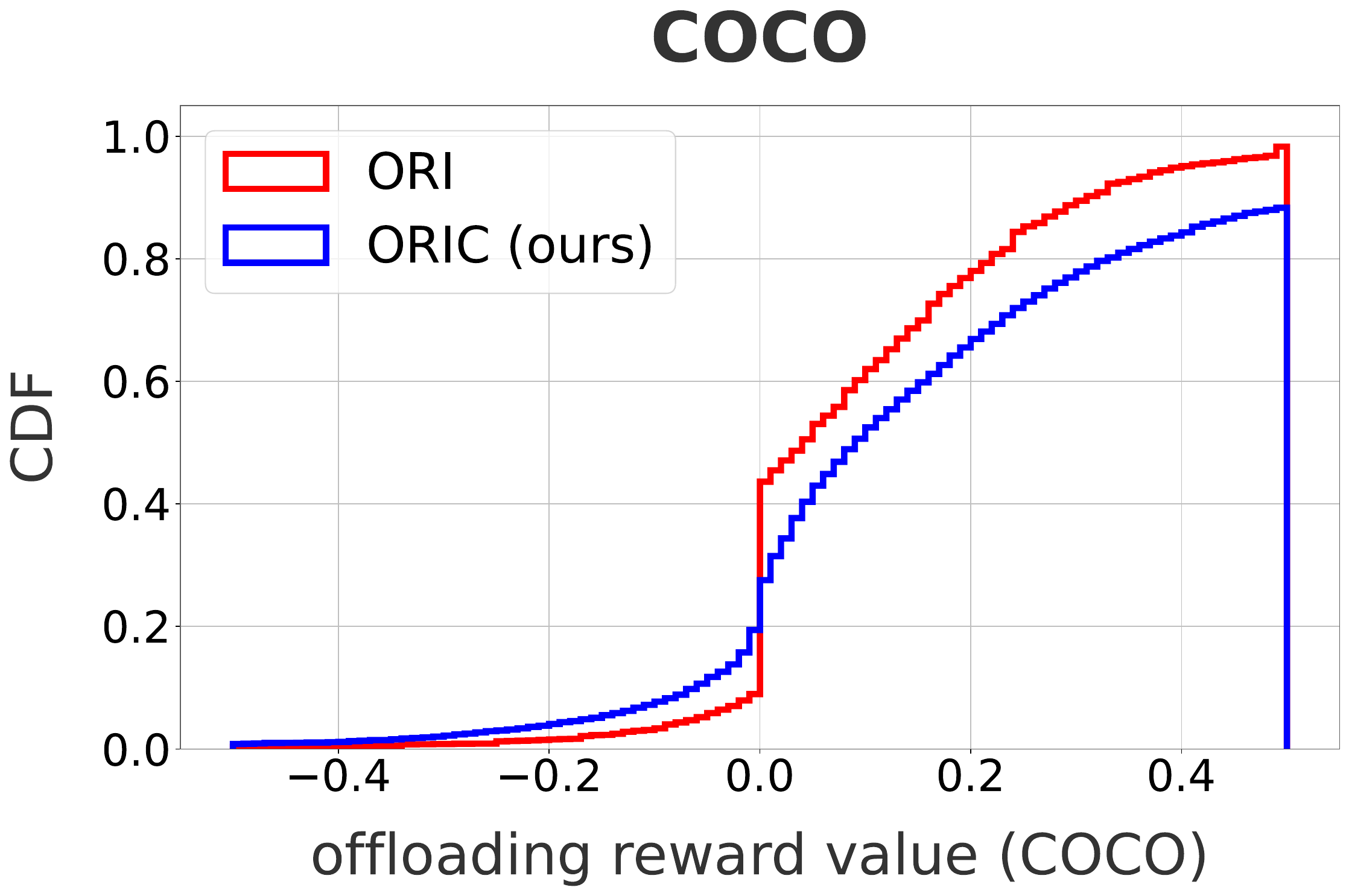}
    \includegraphics[width=0.49\linewidth]{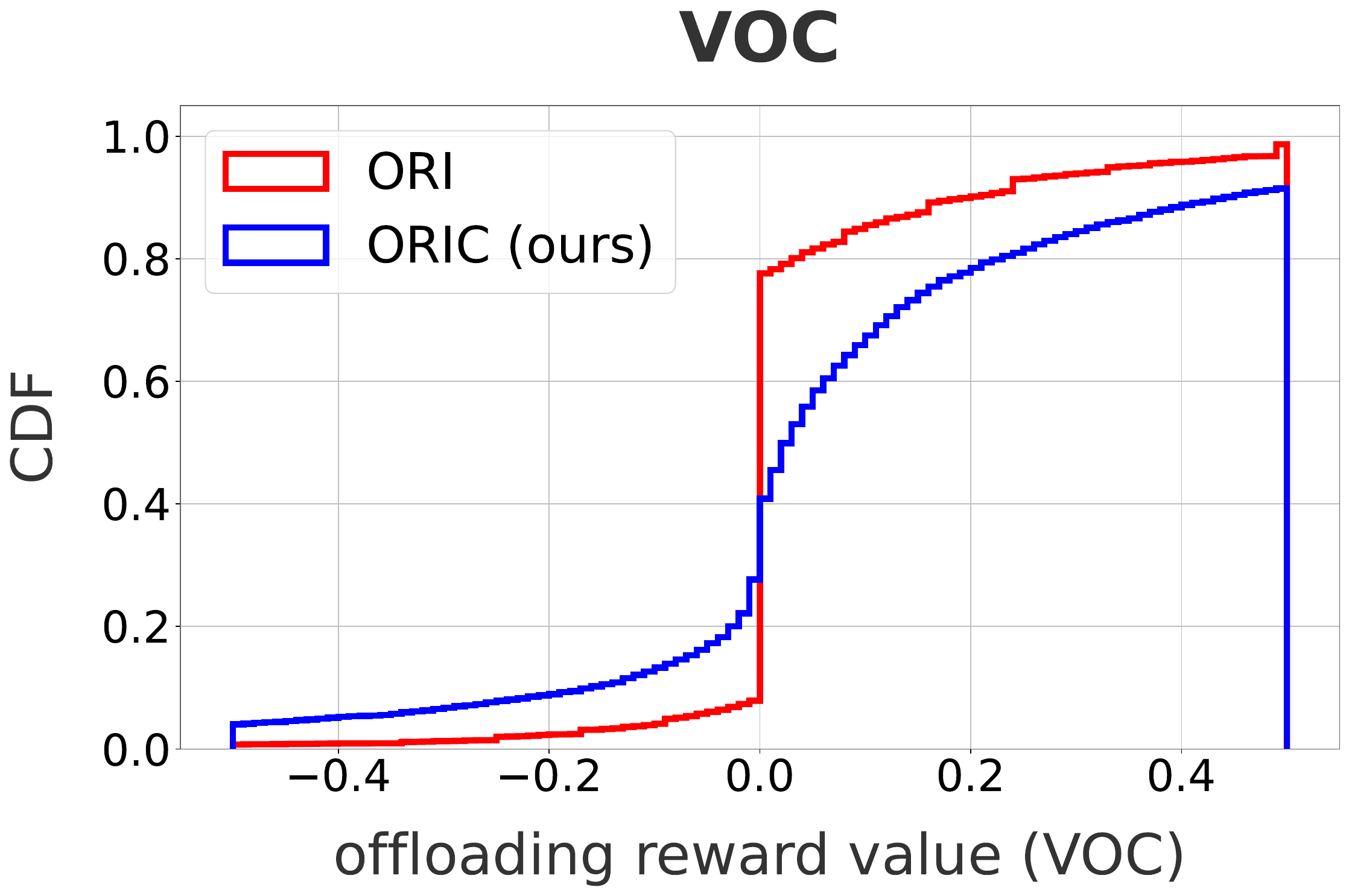}
    \caption{Distribution of ORIC and ORI on COCO (Left) and VOC (Right) with YOLOv5n and YOLOv5m being the object detector pair.}
    \label{fig:oric_dist}
\end{figure}

Specifically, \fig{fig:oric_dist} illustrates the cumulative distribution functions (CDFs) of both ORIC and ORI across the COCO and VOC training datasets, using YOLOv5n and YOLOv5m as the detector pair. The relatively large increase around $0$ for both distributions suggests that many images have an offloading reward close to $0$, \ie both detectors perform similarly. This is intuitive and consistent with observations made by others~\cite{zhou2017adaptive}. The inclusion of the representative image set $\mathcal{E}$ in the computation of ORIC makes this transition somewhat smoother than with ORI, but the presence of a relatively sharp transition remains. 

This can make reward estimation challenging. In particular, an estimator attempting to predict ORIC (or ORI) might achieve a low Mean-Squared Error (MSE) by predominantly predicting near-zero $0$ values.  This could make offloading decisions fragile, with small differences in reward estimates translating in different decisions. To address this issue, prior to training our estimator, we transform the distribution of ORIC to produce a uniform spread of values in $[0,1]$. This is readily achieved by mapping an ORIC value to its position in its underlying CDF. The resulting modified ORIC values (MORIC) are as follows:
\begin{equation}
\label{eq:moric}
    MORIC_i = \cdf(ORIC_i).
\end{equation}
In other words, MORIC values now represent a normalized offloading reward \emph{rank}. 

Next, the training of our estimator proceeds using a weighted MSE loss $\mathcal{L}$ of the form:
\begin{equation}
    \mathcal{L} = \sum_{i=1}^N MORIC_i\cdot \big[e(h_{i, w}) - MORIC_i\big]^2.
\end{equation}
As $MORIC_i$ is in $[0,1]$, this emphasizes errors in high-reward images when computing reward estimates.  The motivations for such a choice are two-fold.  First, most practical edge computing systems operate with low offloading ratios, typically well below 50\%~\cite{nalaie2022deepscale, ghosh2023react}.  This calls for focusing attention on ensuring higher accuracy for high-reward images.  Second, as Table~\ref{tab:image_class_map} illustrated, the set of images for which the weak detector provides more accurate results (negative reward or low rank) is not only small, but also, offloading those images, while wasteful, only produces a small degradation in mAP. 

\begin{figure}[t]
    \centering
    \includegraphics[width=0.8\linewidth]{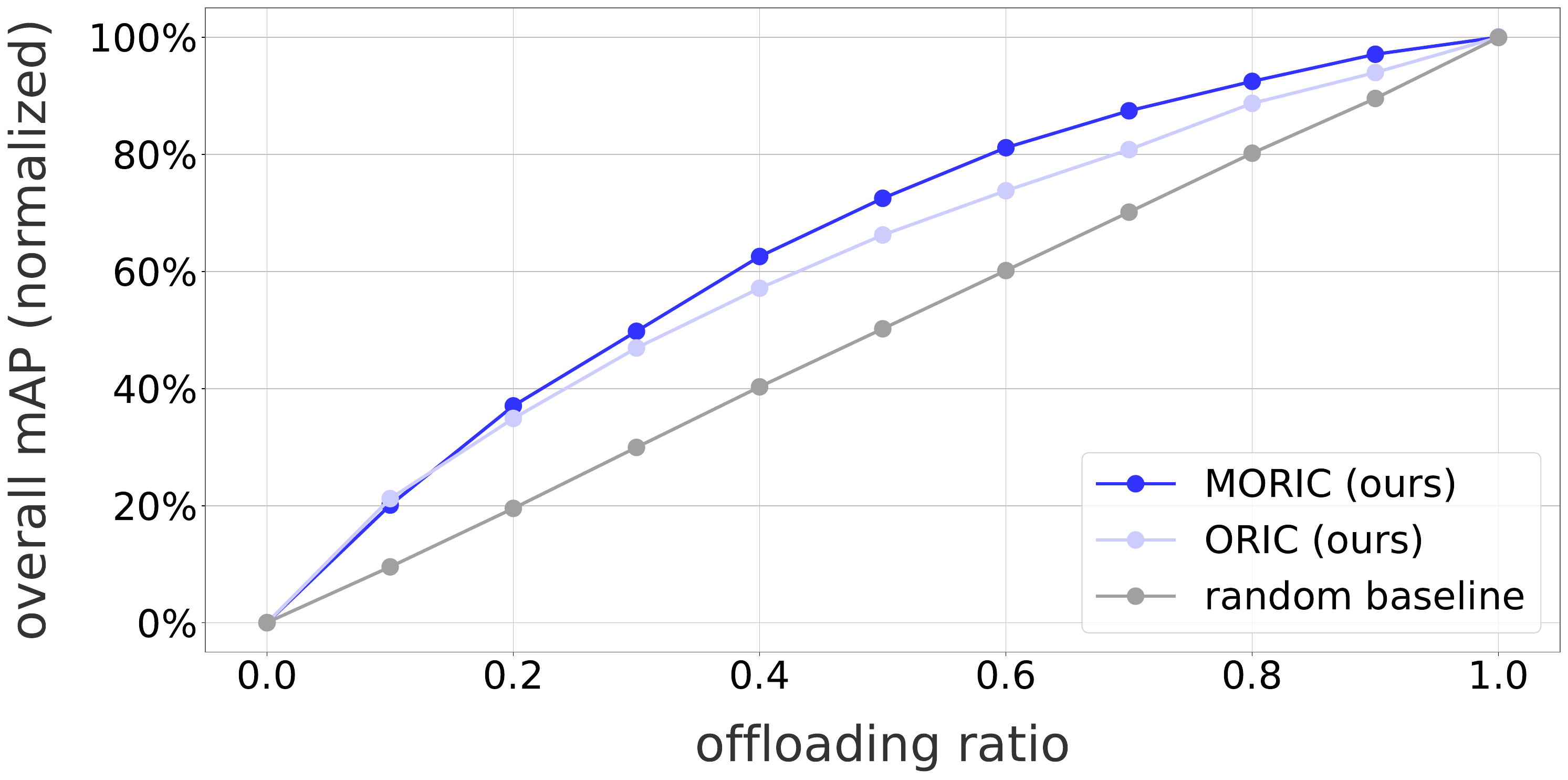}
    \caption{MORIC vs.~ORIC's offloading performance.}
    \label{fig:oric_modification}
\end{figure}

\fig{fig:oric_modification} illustrates the efficacy of the proposed estimation procedure by comparing it to a baseline that offloads images randomly, and a ``vanilla'' estimation procedure that omits to normalize ORIC's distribution and weighs all errors equally in our linear regression.  The results are obtained using again the COCO image dataset with YOLOv5n and YOLOv5m as the object detector pair. The reported mAP was normalized so that values of $0$\% and $100$\% correspond to the mAPs of the weak and strong detectors, respectively. This normalization facilitates comparing performance across offloading ratios. 

\subsubsection*{Observations} Of note in the results of \fig{fig:oric_modification} is the fact that improvements in mAP remain a monotonic function of the offloading ratio.  This is somewhat disappointing as the data from Table~\ref{tab:image_class_map} and \fig{fig:map_representation_size} indicate that it is possible to outperform the strong detector by judiciously relying on the weak detector's output for some images.  Clearly, our reward estimator is unable to pinpoint those images, even when the offloading ratio is not the limiting factor.

In hindsight, this is not overly surprising given that our estimator has access to neither the strong detector's results, nor ground truth annotations.  Nevertheless, and in spite of its limitations, our estimation procedure is capable of outperforming (by over $60\%$ for $r=0.4$) the blind offloading decisions of the random baseline.  In other words, while we are unable to precisely capture the ``best'' images to offload, we are still able to tease apart images with low and high rewards and outperform random selection.

\subsection{ORIC vs.~other Metrics}
\label{sec:oric_estimate}

\begin{figure}[t]
    \centering
    \includegraphics[width=0.8\linewidth]{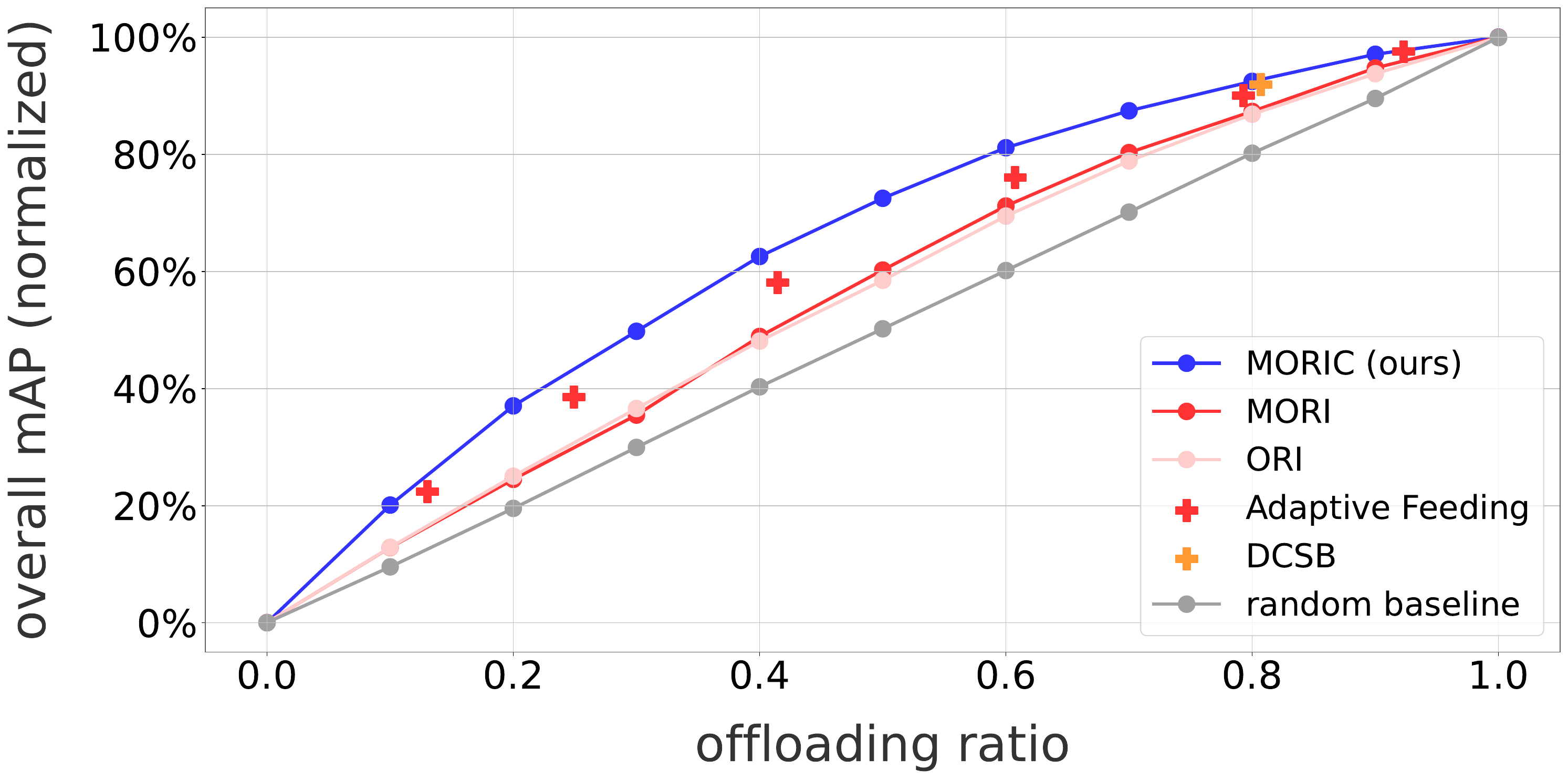}
    \caption{Offloading performance for different reward metrics and estimates.}
    \label{fig:oric_baseline}
\end{figure}

We next compare our MORIC-based estimation approach with several possible alternatives for computing an offloading reward.  They include:
\begin{itemize}[leftmargin=*,nosep]
    \item ORI: This is assuming that ORI rather than ORIC is used as the offloading reward metric. As ORI also spans a continuous range of values, a regression model in the form of an MLP was trained to serve as its estimator.
    \item MORI: Based on ORI, but with similar adjustments as in MORIC, \ie normalized distribution and weighted loss function, when training the regression model as an estimator.
    \item Adaptive Feeding: This is the approach proposed in~\cite{zhou2017adaptive}, where image offloading decisions are cast as a binary classification problem. Images are easy when they have zero or negative ORI values, and difficult otherwise. Only the latter are offloaded. A Support Vector Machine (SVM) classifier is trained using the weak detector's results to distinguish easy from difficult images. 
    
    A weighting factor $c_{+1}$ is used
    to adjust how many images are predicted as difficult.  We used a range of $c_{+1}$ values from $2^{-3}$ to $2^2$ to evaluate the performance of this approach for different offloading ratios.  We note that $c_{+1}$ must be specified at training time and requires iterations to converge to a target offloading ratio.  In contrast, the offloading ratio of our approach can be selected at runtime.
    \item DCSB: This is the approach proposed in~\cite{cao2023edge}. It is also formulated as a binary classification problem, now based on whether the strong detector is expected to predict more objects than the weak detector. It relies on a rule-based policy in making offloading decisions. 
    
    Specifically, it predicts whether the strong detector detects more objects than the weak detector (the image is offloaded) by applying thresholds on two signals from the weak detector's output: the number of objects and the smallest bounding box area detected. The thresholds are chosen to maximize prediction accuracy during training. Note that this maximization produces a unique set of thresholds that then forces the offloading ratio to a fixed value.
\end{itemize}

The comparison was again carried out over the validation sets from COCO (5,000 images) and VOC (4,952 images), with YOLOv5n and YOLOv5m as the object detector pair. A 5-fold cross-validation was used on both datasets to generate the mAP of each alternative. 
The results are in~\fig{fig:oric_baseline}, which demonstrates that our MORIC-based approach consistently achieves the highest mAP at all offloading ratios. 

Of the three approaches that rely on ORI as their offloading reward metric, Adaptive Feeding~\cite{zhou2017adaptive} yields a better offloading performance. As mentioned, this is, however, at the cost of less flexibility in adapting to different offloading ratios. The desired offloading ratio must be specified before training and used to iteratively (binary search) find the correct value for $c_{+1}$.  DCSB, which is also classification-based, exhibits an even more severe limitation.  It performs nearly as well as we do, but its training produces a fixed offloading ratio of $80\%$, which would be infeasible in most edge offloading scenarios.

\begin{figure*}[t]
    \centering
    \includegraphics[width=0.8\linewidth]{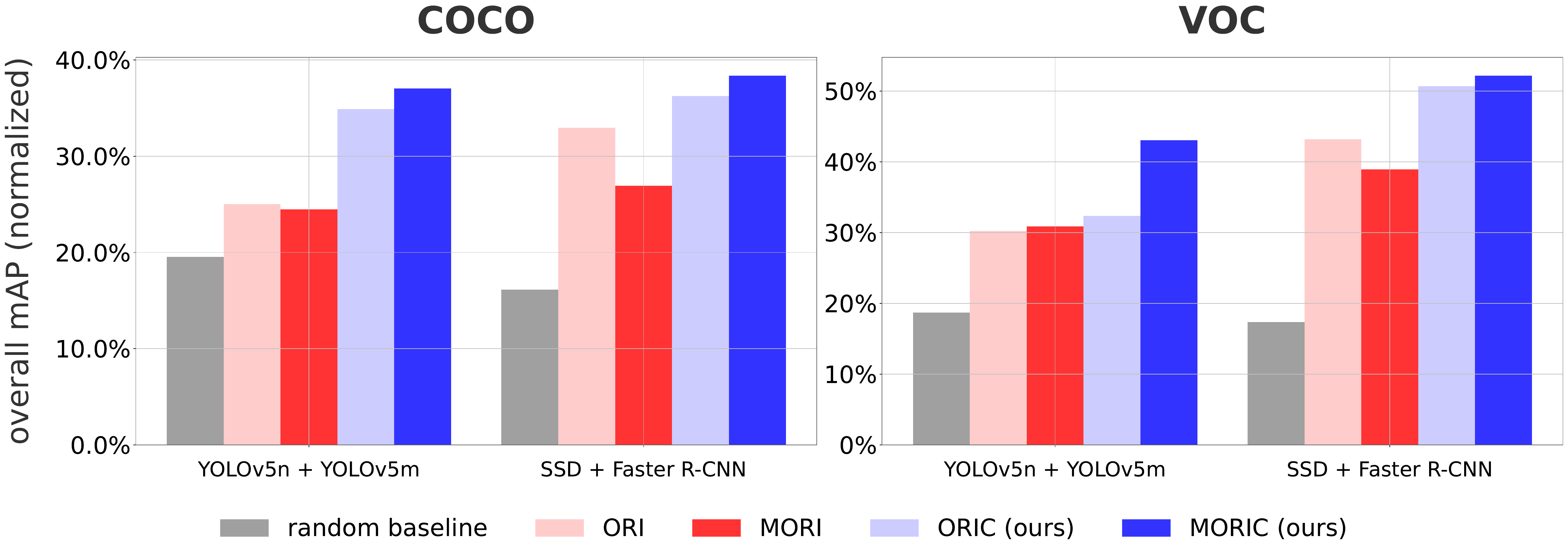}
    \caption{Overall (normalized) mAP for different detector pairs (YOLOv5n+YOLOv5m and SSD+Faster R-CNN) on different datasets (COCO, VOC) and with an offloading ratio of $0.1$.}
    \label{fig:model_generality}
\end{figure*}

As our approach (MORIC) and the one based on estimating ORI as an alternate reward metrics are the only two that readily adapt to different offloading ratios, we explore their relative performance across additional configurations.  Those configurations are intended to assess the robustness of our findings across not only different datasets (COCO and VOC), but also different pairs of weak and strong detectors.

To that end, we conducted evaluations using the COCO and VOC datasets, and two pairs of detectors, the YOLOv5n and YOLOv5m pair on which we previously relied, and a pair of detectors consisting of SSD~\cite{ssd} with a MobileNetV3 Large backbone\footnote{\url{https://pytorch.org/vision/stable/models/generated/torchvision.models.detection.ssdlite320_mobilenet_v3_large.html}} as the weak detector, and Faster R-CNN~\cite{NIPS2015_14bfa6bb} with a ResNet-50-FPN backbone\footnote{\url{https://pytorch.org/vision/stable/models/generated/torchvision.models.detection.fasterrcnn_resnet50_fpn_v2.html}} as the strong detector. \highlight{\useColor}{There were two reasons for choosing those two detectors pairs.  First, both are widely used thanks to their remarkable efficiency and detection accuracy. Second, while an evaluation based on only two pairs of detectors may not allow us to claim full generality, those two pairs rely on sufficiently different architectures that they provide reasonable evidence that the benefits of ORIC would hold for other combinations of detectors.}

\fig{fig:model_generality} presents the result of these additional evaluations for an offloading ratio of $0.2$ representative of many edge offloading deployments.  ORIC-based approaches again systematically outperform ORI-based solutions, with a maximum gain in mAP (over no offloading) of $52.2$\% achieved on VOC using SSD and Faster R-CNN as the detector pair. The gap between MORIC and ORIC is modest but consistent across scenarios. The same does not hold for ORI and MORI.  This is likely because, as shown in~\fig{fig:oric_dist}, the single-image focus of ORI results in a significant percentage of images with a reward of exactly $0$. In such a scenario, the normalization method of \Eqref{eq:moric} fails to ``spread-out'' images with a reward of~$0$.

\section{Experimental Evaluation}
\label{sec:evaluation}

We implemented an edge computing testbed to evaluate the detection performance of our MORIC-based estimation approach in a realistic setting.  The testbed also allowed us to verify that the overhead of implementing our solution was minimal in relation to the full object detection pipeline.

\subsubsection{System setup}
As most object detection applications deployed in an edge computing setting have real-time requirements, we chose the \emph{Nvidia Jetson Xavier NX} (Jetson) as the embedded device in our testbed. The Jetson platform boasts a 6-core NVIDIA Carmel ARM CPU, a 384-core NVIDIA Volta GPU with 48 Tensor Cores, and a unified 8GB RAM shared by the CPU and GPU. This ensures that we can run relatively accurate (weak) detectors on our embedded device without incurring unreasonable processing latency. Our edge server consisted of an 8-core Intel(R) Core(TM) i7-10700K CPU @ 3.80GHz and an Nvidia GeForce RTX 3090 GPU. Data transmission following offloading decisions is realized by a direct connection using a 1-Gbps Ethernet cable linking the local device to the edge server.

To optimize on-board inference speed for both our weak object detectors and the MORIC estimation model, we leveraged TensorRT. This optimization was applied through the built-in NVIDIA CUDA parallel programming model and floating point 16 (FP16) optimizations available on the Jetson platform used on the local device.

\subsubsection{Runtime Evaluation}

\begin{table*}[t]
\renewcommand{\arraystretch}{1.3}
\begin{center}
\caption{Relative contributions of the different parts of the object detection pipeline}
\label{tab:pipeline_time}
\begin{tabular}{|c|ccc|ccc|}
\hline
& \multicolumn{3}{c|}{\textbf{YOLOv5n + YOLOv5m }} & \multicolumn{3}{c|}{\textbf{SSD + Faster R-CNN}} \\
\hline
\multirow{2}{*}{\textbf{Component}} & \textbf{Absolute} & \multicolumn{2}{c|}{\textbf{Relative}} & \textbf{Absolute} & \multicolumn{2}{c|}{\textbf{Relative}}\\
& \textbf{mean(std)} (ms) & \textbf{not offloaded} & \textbf{offloaded} & \textbf{mean(std)} (ms) & \textbf{not offloaded} & \textbf{offloaded}\\
\hline
Weak Detector & $8.92(0.16)$ & $94.41\%$ & $21.77\%$ & $5.29(0.14)$ & $90.18\%$ & $7.70\%$\\
\hline
ORIC Estimation & $0.53(0.04)$ & $5.59\%$ & $1.35\%$ & $0.57(0.06)$ & $9.82\%$ & $0.83\%$\\
\hline
Transmission & $19.24(7.44)$ & $-$ & $46.96\%$ & $18.30(6.56)$ & $-$ & $26.65\%$\\
\hline
Strong Detector & $12.26(8.45)$ & $-$ & $29.92\%$ & $44.50(32.43)$ & $-$ & $64.82\%$\\
\hline
\end{tabular}
\end{center}
\end{table*}

\begin{figure*}[t]
    \centering
    \includegraphics[width=0.8\linewidth]{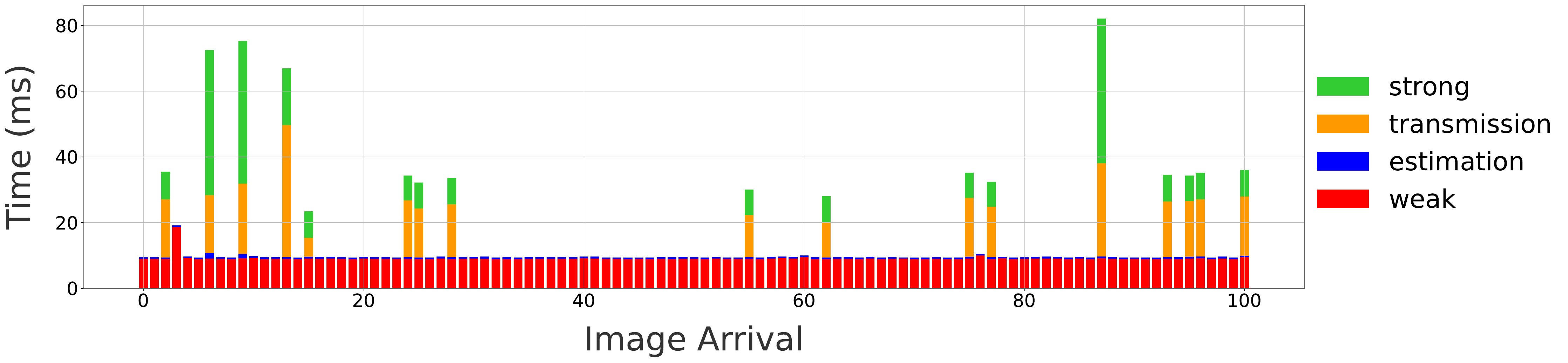}
    \caption{Break-down of the object detection pipeline into weak detection time, reward estimation time, image transmission time and strong detection time for a sample trace of $100$ images with an offloading ratio of $0.2$, using YOLOv5n and YOLOv5m as the weak and strong detector, respectively.}
    \label{fig:detection_pipeline}
\end{figure*}

To ensure a comprehensive examination of the runtime computational overhead and offloading efficacy of our reward estimation approach, we considered the two object detector pairs previously discussed in Section~\ref{sec:oric_estimate}: namely, YOLOv5n and YOLOv5m, as well as SSD and Faster R-CNN. The COCO dataset served as the benchmark for our runtime evaluation.

For each image, our edge object detection pipeline has up to four parts: (i) weak detector's inference, (ii) reward estimation, (iii) image transmission, and (iv) strong detector's inference.  The latter two are present only if an image is offloaded. We report statistics of the absolute and relative time spent in each part in Table~\ref{tab:pipeline_time}, and a visualization of the detection pipeline in~\fig{fig:detection_pipeline} for a representative sample trace of $100$ images. 

As Table~\ref{tab:pipeline_time} indicates, on the Jetson platform, reward estimation introduces a minimal overhead of about $0.5$ms. This accounts for only $1.35$\% ($0.83$\%) and $5.59$\% ($9.82$\%) of the total detection pipeline latency for images that are offloaded and not offloaded, respectively, when deployed on  YOLOv5n and YOLOv5m (SSD and Faster R-CNN) as the detector pair. Note that image transmission and strong detection times are much more variable than local detection and estimation times. This is because we fixed the image size for the TensorRT-accelerated weak detectors ($640\times640$ for YOLOv5n and $320\times320$ for SSD), while allowing the original images whose sizes vary to be offloaded to the strong detector. 

We next extend our investigation to exploring the trade-off between total detection time and detection accuracy as a function of the offloading ratio.

\begin{figure}[t]
    \centering
    \includegraphics[width=\linewidth]{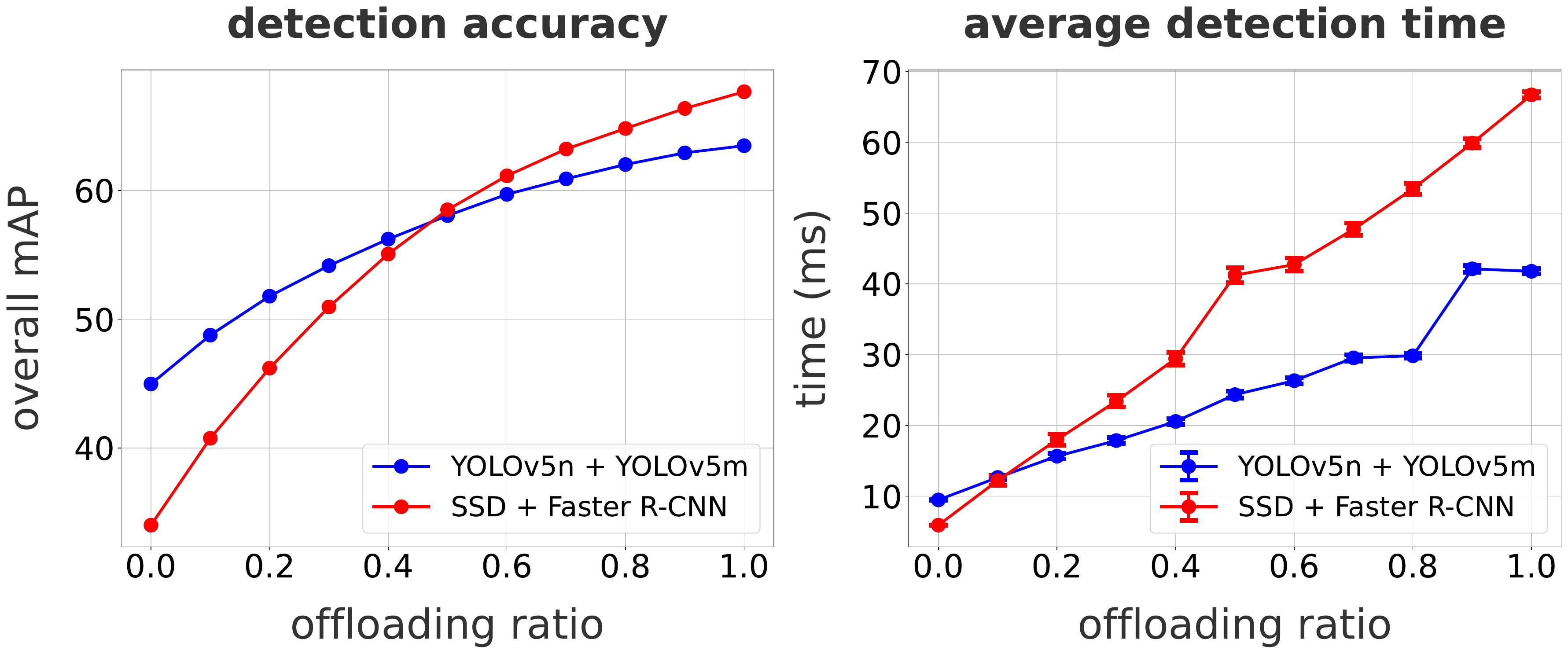}
    \caption{Detection accuracy (Left) and average detection time (Right) of two object detector pairs (YOLOv5n \& YOLOv5m, SSD \& Faster R-CNN) in relation to the offloading ratio.}
    \label{fig:testbed}
\end{figure}

\fig{fig:testbed} presents the mAP values achieved by our edge computing testbed and the average and $95$\% confidence interval of the total detection time, both as a function of the offloading ratio. 
As expected, total detection latency increases as more images are offloaded. In our single local device testbed, overall detection time is a mostly linear function of the offloading ratio.  This may, however, not hold in a general multi-device scenario, where uncoordinated transmissions across local devices may cause network congestion or resource overload at the edge server.  Those may increase non-linearly with the offloading ratio, \eg as shown in~\cite{yoon2021implementing}.  From that standpoint, the fact that our offloading solution shows faster increases in mAP for small values of the offloading ratio (the functions are concave), highlights that it can yield meaningful benefits while preserving most of the advantages of a distributed solution.

\section{Related Work}
\label{sec:related_work}

\subsection{Recent Advances in Object Detection}

Over the past decade, deep neural networks (DNNs) have become essential in developing increasingly fast and accurate object detectors. Existing models can be mainly categorized as two-stage (e.g., Faster R-CNN~\cite{girshick2015fast, NIPS2015_14bfa6bb, He_2017_ICCV} and R-FCN~\cite{dai2016r}) and one-stage detectors (e.g., YOLO~\cite{redmon2016you, redmon2017yolo9000, yolov3} and SSD~\cite{ssd}), with the latter offering faster detection by predicting bounding boxes and class labels simultaneously. YOLO’s balance between accuracy and efficiency has made it popular, especially for real-time applications~\cite{terven2023comprehensive}.

An approach, presented in~\cite{zhou2017adaptive}, introduces an adaptive strategy that seeks to combine the strengths of both fast and more accurate models. Each image chooses between a fast and a more accurate model for detection by performing a binary classification using SVM based on its ORI value estimate. Although it does not explicitly target edge computing, it is closely related to our work and is the method behind the ``Adaptive Feeding'' solution to which we compare our approach in Section~\ref{sec:oric_estimate}.

Other advances~\cite{zhang2021csl, kim2020method} explored approaches to reduce the computational overhead of object detection by proposing novel computational techniques or exploiting hardware characteristics.  Such solutions further contribute to making object detectors suitable for edge computing deployment.

\subsection{Edge AI Techniques}
Edge computing has attracted substantial attention within the domain of AI applications, as it provides an effective mechanism to circumvent the constraints of limited computational resources by offloading part of the inference tasks. Pertinent literature on the distribution of computational workloads to the edge can be broadly categorized into two paradigms~\cite{matsubara2022split}.

The first paradigm, split computing, segments the workload by partitioning large DNN models into head and tail models that are, executed on local devices and edge servers, respectively. Early works~\cite{kang2017neurosurgeon, eshratifar2019jointdnn} explore optimal partitioning strategies that mitigate local computation and network transmission overheads. Subsequent contributions~\cite{eshratifar2019bottlenet, shao2020bottlenet++, matsubara2019distilled, matsubara2020head} consider using autoencoders to further compact the head models. Such strategies facilitate minimized data transmissions by offloading the compressed intermediate computational results. More recently, dropout techniques have been introduced as part of autoencoders training to tackle bandwidth fluctuation or data loss during task offloading. This allows progressively more accurate reconstruction at edge servers as more data is received~\cite{hu2020starfish, wang2023progressive}. These works, however, require combining both local and edge computations to complete inference tasks.

Another paradigm
known as ``model cascade'' and/or ``early exiting'', \highlight{\useColor}{and the one this paper is based on,} takes a different approach. It empowers local devices to generate the final inference output, with tasks offloaded to the edge only when local inference outcomes are deemed inadequate. The model cascade approach employs a synergistic combination of a weak model at the local device and a strong model at the edge server~\cite{wang2017idk}, with \cite{chakrabarti2021real, qiu2022adaptive} extending it by constraining offloading decisions through a rate control mechanism to avoid transmission bursts that might overload the network or the edge servers.  Early exiting~\cite{teerapittayanon2016branchynet, teerapittayanon2017distributed, laskaridis2020spinn, zeng2019boomerang} focuses on avoiding redundant and unnecessary DNN inference by integrating sub-branches to the partition points \highlight{\useColor}{(between local devices and the edge)} of the DNN models. \highlight{\useColor}{The system can exit early at a sub-branch if the intermediate result from local inference exhibits sufficient confidence,} permitting a retreat from offloading. 

\highlight{\useColor}{Our approach applies to both model cascade and early exiting models.  This is because, as discussed in Section~\ref{sec:oric_alt}, predicting the mAP gain from offloading using our ORIC-based estimation approach can be realized using either the output of the object detector (cascade), or the feature maps from its intermediate hidden layers (early exiting).}

An important difference between the paper and these works is that most focus on image classification and few consider object detection. This greatly simplifies offloading decisions as
classification entropy can readily be used as a confidence indicator to evaluate the benefit of offloading.  A comparable metric is largely absent for object detection, which was our primary focus in developing ORIC (and MORIC). 

\subsection{Edge Object Detection}
Many studies have investigated edge/cloud computing to assist video analytics applications that rely on object detection. 

Early works, faced with limited on-board computational resources, relied entirely on edge servers for object detection inferences, with local devices only handling lightweight computational tasks such as object tracking~\cite{chen2015glimpse}, Region-of-Interest (RoI) encoding~\cite{liu2019edge}, and frame size adaptation~\cite{liu2018dare}. 

More recent works have benefited from advances in both object detection model architectures and computational hardware, and have increasingly contemplated a hybrid paradigm involving both local and edge inferences. To reduce inference and offloading overhead, DeepScale~\cite{nalaie2022deepscale} suggests dynamic video frame size adjustments, ensuring only minimal degradation on detection accuracy. This approach reveals a trade-off between accuracy and latency compared to both edge-only or local-only alternatives. React~\cite{ghosh2023react} shares our study's perspective that edge and local detection can be complementary.  It proposes an algorithm that fuses edge and local detection results to achieve a high mAP with reduced latency. We note that while collaborative approaches involving both local devices and the edge are gaining traction, most proposals to-date rely on rudimentary offloading decision-making strategies, \eg offloading once every $K$ frames. As a result, most would benefit from our approach to help them more accurately select frames that would most benefit from offloading.

A few studies have focused on optimizing offloading decisions in edge object detection systems. \cite{yoon2021implementing} considers a multi-device environment, where each device chooses between local execution and offloading to maximize the overall inference speed of object detection.  The goal is to avoid inefficient resource utilization that arises from competition among local devices. This approach deploys identical object detection models on both the local devices and the edge server, thereby neglecting any potential enhancement in detection accuracy via offloading. \cite{ku2021adaptive} investigates the optimal partitioning and offloading of object detection tasks in a vehicular edge computing scenario, targeting both detection accuracy maximization and inference latency minimization. However, its accuracy optimization hinges on adjusting image quality based on real-time channel conditions and local energy constraints.

Last but not least, \cite{cao2023edge} aligns closely with our  optimization objectives and is the method behind the DCSB solution we explored in Section~\ref{sec:oric_estimate}. However, and as discussed earlier, it formulates offloading as a binary classification problem and proposes a rule-based approach that yields a fixed offloading ratio, \eg $80\%$ as illustrated in \fig{fig:oric_baseline}. This limits its applicability in the context of edge offloading applications.

\section{Conclusion}
\label{sec:conclusion}

This paper proposes ORIC, an offloading reward metric designed to quantify the enhancement in the detection accuracy metric mAP, from offloading an individual image to a stronger edge object detector. ORIC was shown to improve on existing metrics with a similar goal, primarily because it incorporates the overall context on which mAP relies to measure detection accuracy. The paper also develops a practical approach to efficiently estimate ORIC using only the output of the local (weak) detector. The benefits of the approach were evaluated across configurations involving different image datasets and combinations of weak and strong detectors, as well as on an experimental testbed implemented to emulate a realistic edge computing setting. The evaluation showed that our solution outperforms existing alternatives, with some configurations allowing it to extract over $50\%$ of the strong detector's benefits with an offloading ratio of just $20\%$.

We note that while ORIC's estimation assumed no correlation between successive images, accommodating correlation can be realized with existing solutions. For example, as in~\cite{qiu2022adaptive}, a Deep Q-Network could be added and fed ORIC estimates of the last $k$ images to make offloading decisions that would now account for possible correlation across images.  
Similarly, while, for simplicity, ORIC's estimation relied directly on the output of the weak detector, other options are possible, \eg feature maps from the weak detector's hidden layers.  This could facilitate integrating ORIC with early exit strategies commonly used in edge computing settings.
Finally, we anticipate that ORIC can be incorporated into edge optimization frameworks~\cite{chen2018optimized, chakrabarti2021real} that target system performance metrics such as latency or bandwidth, and extend their applicability to object detection settings.

\clearpage
\bibliographystyle{IEEEtran}
\bibliography{reference}

\begin{thebibliography}{10}
\providecommand{\url}[1]{#1}
\csname url@samestyle\endcsname
\providecommand{\newblock}{\relax}
\providecommand{\bibinfo}[2]{#2}
\providecommand{\BIBentrySTDinterwordspacing}{\spaceskip=0pt\relax}
\providecommand{\BIBentryALTinterwordstretchfactor}{4}
\providecommand{\BIBentryALTinterwordspacing}{\spaceskip=\fontdimen2\font plus
\BIBentryALTinterwordstretchfactor\fontdimen3\font minus \fontdimen4\font\relax}
\providecommand{\BIBforeignlanguage}[2]{{%
\expandafter\ifx\csname l@#1\endcsname\relax
\typeout{** WARNING: IEEEtran.bst: No hyphenation pattern has been}%
\typeout{** loaded for the language `#1'. Using the pattern for}%
\typeout{** the default language instead.}%
\else
\language=\csname l@#1\endcsname
\fi
#2}}
\providecommand{\BIBdecl}{\relax}
\BIBdecl

\bibitem{NIPS2015_14bfa6bb}
S.~Ren, K.~He, R.~Girshick, and J.~Sun, ``Faster r-cnn: Towards real-time object detection with region proposal networks,'' in \emph{Advances in Neural Information Processing Systems}, vol.~28.\hskip 1em plus 0.5em minus 0.4em\relax Curran Associates, Inc., 2015.

\bibitem{He_2017_ICCV}
K.~He, G.~Gkioxari, P.~Dollar, and R.~Girshick, ``Mask r-cnn,'' in \emph{Proceedings of the IEEE International Conference on Computer Vision (ICCV)}, Oct 2017.

\bibitem{Lin_2017_CVPR}
T.-Y. Lin, P.~Dollar, R.~Girshick, K.~He, B.~Hariharan, and S.~Belongie, ``Feature pyramid networks for object detection,'' in \emph{Proceedings of the IEEE Conference on Computer Vision and Pattern Recognition (CVPR)}, July 2017.

\bibitem{yolov3}
\BIBentryALTinterwordspacing
J.~Redmon and A.~Farhadi, ``Yolov3: An incremental improvement,'' 2018. [Online]. Available: \url{https://arxiv.org/abs/1804.02767}
\BIBentrySTDinterwordspacing

\bibitem{ssd}
W.~Liu, D.~Anguelov, D.~Erhan, C.~Szegedy, S.~Reed, C.-Y. Fu, and A.~C. Berg, ``Ssd: Single shot multibox detector,'' in \emph{Computer Vision -- ECCV 2016}.\hskip 1em plus 0.5em minus 0.4em\relax Springer International Publishing, 2016, pp. 21--37.

\bibitem{Tan_2020_CVPR}
M.~Tan, R.~Pang, and Q.~V. Le, ``Efficientdet: Scalable and efficient object detection,'' in \emph{Proceedings of the IEEE/CVF Conference on Computer Vision and Pattern Recognition (CVPR)}, June 2020.

\bibitem{kim2021deep}
J.-H. Kim, N.~Kim, and C.~S. Won, ``Deep edge computing for videos,'' \emph{IEEE Access}, vol.~9, pp. 123\,348--123\,357, 2021.

\bibitem{nalaie2022deepscale}
K.~Nalaie, R.~Xu, and R.~Zheng, ``Deepscale: Online frame size adaptation for multi-object tracking on smart cameras and edge servers,'' in \emph{2022 IEEE/ACM Seventh International Conference on Internet-of-Things Design and Implementation (IoTDI)}.\hskip 1em plus 0.5em minus 0.4em\relax IEEE, 2022, pp. 67--79.

\bibitem{ghosh2023react}
A.~Ghosh, S.~Iyengar, S.~Lee, A.~Rathore, and V.~N. Padmanabhan, ``React: Streaming video analytics on the edge with asynchronous cloud support,'' in \emph{Proceedings of the 8th ACM/IEEE Conference on Internet of Things Design and Implementation}, 2023, pp. 222--235.

\bibitem{tan2021deep}
T.~Tan and G.~Cao, ``Deep learning video analytics on edge computing devices,'' in \emph{2021 18th Annual IEEE International Conference on Sensing, Communication, and Networking (SECON)}.\hskip 1em plus 0.5em minus 0.4em\relax IEEE, 2021, pp. 1--9.

\bibitem{hanyao2021edge}
M.~Hanyao, Y.~Jin, Z.~Qian, S.~Zhang, and S.~Lu, ``Edge-assisted online on-device object detection for real-time video analytics,'' in \emph{IEEE INFOCOM 2021-IEEE Conference on Computer Communications}.\hskip 1em plus 0.5em minus 0.4em\relax IEEE, 2021, pp. 1--10.

\bibitem{deng2020fedvision}
Y.~Deng, T.~Han, and N.~Ansari, ``Fedvision: Federated video analytics with edge computing,'' \emph{IEEE Open Journal of the Computer Society}, vol.~1, pp. 62--72, 2020.

\bibitem{zhou2017adaptive}
H.-Y. Zhou, B.-B. Gao, and J.~Wu, ``Adaptive feeding: Achieving fast and accurate detections by adaptively combining object detectors,'' in \emph{Proceedings of the IEEE International Conference on Computer Vision}, 2017, pp. 3505--3513.

\bibitem{cao2023edge}
Z.~Cao, Z.~Li, Y.~Chen, H.~Pan, Y.~Hu, and J.~Liu, ``Edge-cloud collaborated object detection via difficult-case discriminator,'' in \emph{2023 IEEE 43rd International Conference on Distributed Computing Systems (ICDCS)}.\hskip 1em plus 0.5em minus 0.4em\relax IEEE, 2023, pp. 259--270.

\bibitem{padilla20}
R.~Padilla, S.~L. Netto, and E.~A.~B. da~Silva, ``A survey on performance metrics for object-detection algorithms,'' in \emph{2020 International Conference on Systems, Signals and Image Processing (IWSSIP)}, 2020, pp. 237--242.

\bibitem{medium2018map}
\BIBentryALTinterwordspacing
(2018) map (mean average precision) for object detection. [Online]. Available: \url{https://jonathan-hui.medium.com/map-mean-average-precision-for-object-detection-45c121a31173}
\BIBentrySTDinterwordspacing

\bibitem{teerapittayanon2016branchynet}
S.~Teerapittayanon, B.~McDanel, and H.-T. Kung, ``Branchynet: Fast inference via early exiting from deep neural networks,'' in \emph{2016 23rd international conference on pattern recognition (ICPR)}.\hskip 1em plus 0.5em minus 0.4em\relax IEEE, 2016, pp. 2464--2469.

\bibitem{teerapittayanon2017distributed}
------, ``Distributed deep neural networks over the cloud, the edge and end devices,'' in \emph{2017 IEEE 37th international conference on distributed computing systems (ICDCS)}.\hskip 1em plus 0.5em minus 0.4em\relax IEEE, 2017, pp. 328--339.

\bibitem{laskaridis2020spinn}
S.~Laskaridis, S.~I. Venieris, M.~Almeida, I.~Leontiadis, and N.~D. Lane, ``Spinn: synergistic progressive inference of neural networks over device and cloud,'' in \emph{Proceedings of the 26th annual international conference on mobile computing and networking}, 2020, pp. 1--15.

\bibitem{zeng2019boomerang}
L.~Zeng, E.~Li, Z.~Zhou, and X.~Chen, ``Boomerang: On-demand cooperative deep neural network inference for edge intelligence on the industrial internet of things,'' \emph{IEEE Network}, vol.~33, no.~5, pp. 96--103, 2019.

\bibitem{lin2014microsoft}
T.-Y. Lin, M.~Maire, S.~Belongie, J.~Hays, P.~Perona, D.~Ramanan, P.~Doll{\'a}r, and C.~L. Zitnick, ``Microsoft coco: Common objects in context,'' in \emph{Computer Vision--ECCV 2014: 13th European Conference, Zurich, Switzerland, September 6-12, 2014, Proceedings, Part V 13}.\hskip 1em plus 0.5em minus 0.4em\relax Springer, 2014, pp. 740--755.

\bibitem{everingham2010pascal}
M.~Everingham, L.~Van~Gool, C.~K. Williams, J.~Winn, and A.~Zisserman, ``The {Pascal} visual object classes {(VOC)} challenge,'' \emph{International journal of computer vision}, vol.~88, pp. 303--338, 2010.

\bibitem{chakrabarti2021real}
A.~Chakrabarti, R.~Gu{\'e}rin, C.~Lu, and J.~Liu, ``Real-time edge classification: Optimal offloading under token bucket constraints,'' in \emph{2021 IEEE/ACM Symposium on Edge Computing (SEC)}.\hskip 1em plus 0.5em minus 0.4em\relax IEEE, 2021, pp. 41--54.

\bibitem{bolya2020tide}
D.~Bolya, S.~Foley, J.~Hays, and J.~Hoffman, ``Tide: A general toolbox for identifying object detection errors,'' in \emph{Computer Vision--ECCV 2020: 16th European Conference, Glasgow, UK, August 23--28, 2020, Proceedings, Part III 16}.\hskip 1em plus 0.5em minus 0.4em\relax Springer, 2020, pp. 558--573.

\bibitem{yoon2021implementing}
G.~Yoon, G.-Y. Kim, H.~Yoo, S.~C. Kim, and R.~Kim, ``Implementing practical dnn-based object detection offloading decision for maximizing detection performance of mobile edge devices,'' \emph{IEEE Access}, vol.~9, pp. 140\,199--140\,211, 2021.

\bibitem{girshick2015fast}
R.~Girshick, ``Fast r-cnn,'' in \emph{Proceedings of the IEEE international conference on computer vision}, 2015, pp. 1440--1448.

\bibitem{dai2016r}
J.~Dai, Y.~Li, K.~He, and J.~Sun, ``R-fcn: Object detection via region-based fully convolutional networks,'' \emph{Advances in neural information processing systems}, vol.~29, 2016.

\bibitem{redmon2016you}
J.~Redmon, S.~Divvala, R.~Girshick, and A.~Farhadi, ``You only look once: Unified, real-time object detection,'' in \emph{Proceedings of the IEEE conference on computer vision and pattern recognition}, 2016, pp. 779--788.

\bibitem{redmon2017yolo9000}
J.~Redmon and A.~Farhadi, ``Yolo9000: better, faster, stronger,'' in \emph{Proceedings of the IEEE conference on computer vision and pattern recognition}, 2017, pp. 7263--7271.

\bibitem{terven2023comprehensive}
J.~Terven and D.~Cordova-Esparza, ``A comprehensive review of yolo: From yolov1 to yolov8 and beyond,'' \emph{arXiv preprint arXiv:2304.00501}, 2023.

\bibitem{zhang2021csl}
Y.-M. Zhang, C.-C. Lee, J.-W. Hsieh, and K.-C. Fan, ``Csl-yolo: A new lightweight object detection system for edge computing,'' \emph{arXiv preprint arXiv:2107.04829}, 2021.

\bibitem{kim2020method}
R.~Kim, G.~Kim, H.~Kim, G.~Yoon, and H.~Yoo, ``A method for optimizing deep learning object detection in edge computing,'' in \emph{2020 International Conference on Information and Communication Technology Convergence (ICTC)}.\hskip 1em plus 0.5em minus 0.4em\relax IEEE, 2020, pp. 1164--1167.

\bibitem{matsubara2022split}
Y.~Matsubara, M.~Levorato, and F.~Restuccia, ``Split computing and early exiting for deep learning applications: Survey and research challenges,'' \emph{ACM Computing Surveys}, vol.~55, no.~5, pp. 1--30, 2022.

\bibitem{kang2017neurosurgeon}
Y.~Kang, J.~Hauswald, C.~Gao, A.~Rovinski, T.~Mudge, J.~Mars, and L.~Tang, ``Neurosurgeon: Collaborative intelligence between the cloud and mobile edge,'' \emph{ACM SIGARCH Computer Architecture News}, vol.~45, no.~1, pp. 615--629, 2017.

\bibitem{eshratifar2019jointdnn}
A.~E. Eshratifar, M.~S. Abrishami, and M.~Pedram, ``Jointdnn: An efficient training and inference engine for intelligent mobile cloud computing services,'' \emph{IEEE Transactions on Mobile Computing}, vol.~20, no.~2, pp. 565--576, 2019.

\bibitem{eshratifar2019bottlenet}
A.~E. Eshratifar, A.~Esmaili, and M.~Pedram, ``Bottlenet: A deep learning architecture for intelligent mobile cloud computing services,'' in \emph{2019 IEEE/ACM International Symposium on Low Power Electronics and Design (ISLPED)}.\hskip 1em plus 0.5em minus 0.4em\relax IEEE, 2019, pp. 1--6.

\bibitem{shao2020bottlenet++}
J.~Shao and J.~Zhang, ``Bottlenet++: An end-to-end approach for feature compression in device-edge co-inference systems,'' in \emph{2020 IEEE International Conference on Communications Workshops (ICC Workshops)}.\hskip 1em plus 0.5em minus 0.4em\relax IEEE, 2020, pp. 1--6.

\bibitem{matsubara2019distilled}
Y.~Matsubara, S.~Baidya, D.~Callegaro, M.~Levorato, and S.~Singh, ``Distilled split deep neural networks for edge-assisted real-time systems,'' in \emph{Proceedings of the 2019 Workshop on Hot Topics in Video Analytics and Intelligent Edges}, 2019, pp. 21--26.

\bibitem{matsubara2020head}
Y.~Matsubara, D.~Callegaro, S.~Baidya, M.~Levorato, and S.~Singh, ``Head network distillation: Splitting distilled deep neural networks for resource-constrained edge computing systems,'' \emph{IEEE Access}, vol.~8, pp. 212\,177--212\,193, 2020.

\bibitem{hu2020starfish}
P.~Hu, J.~Im, Z.~Asgar, and S.~Katti, ``Starfish: Resilient image compression for aiot cameras,'' in \emph{Proceedings of the 18th Conference on Embedded Networked Sensor Systems}, 2020, pp. 395--408.

\bibitem{wang2023progressive}
R.~Wang, H.~Liu, J.~Qiu, M.~Xu, R.~Gu{\'e}rin, and C.~Lu, ``Progressive neural compression for adaptive image offloading under timing constraints,'' in \emph{2023 IEEE Real-Time Systems Symposium (RTSS)}.\hskip 1em plus 0.5em minus 0.4em\relax IEEE, 2023, pp. 118--130.

\bibitem{wang2017idk}
X.~Wang, Y.~Luo, D.~Crankshaw, A.~Tumanov, F.~Yu, and J.~E. Gonzalez, ``Idk cascades: Fast deep learning by learning not to overthink,'' \emph{arXiv preprint arXiv:1706.00885}, 2017.

\bibitem{qiu2022adaptive}
J.~Qiu, R.~Wang, A.~Chakrabarti, R.~Gu{\'e}rin, and C.~Lu, ``Adaptive edge offloading for image classification under rate limit,'' \emph{IEEE Transactions on Computer-Aided Design of Integrated Circuits and Systems}, vol.~41, no.~11, pp. 3886--3897, 2022.

\bibitem{chen2015glimpse}
T.~Y.-H. Chen, L.~Ravindranath, S.~Deng, P.~Bahl, and H.~Balakrishnan, ``Glimpse: Continuous, real-time object recognition on mobile devices,'' in \emph{Proceedings of the 13th ACM Conference on Embedded Networked Sensor Systems}, 2015, pp. 155--168.

\bibitem{liu2019edge}
L.~Liu, H.~Li, and M.~Gruteser, ``Edge assisted real-time object detection for mobile augmented reality,'' in \emph{The 25th annual international conference on mobile computing and networking}, 2019, pp. 1--16.

\bibitem{liu2018dare}
Q.~Liu and T.~Han, ``Dare: Dynamic adaptive mobile augmented reality with edge computing,'' in \emph{2018 IEEE 26th International Conference on Network Protocols (ICNP)}.\hskip 1em plus 0.5em minus 0.4em\relax IEEE, 2018, pp. 1--11.

\bibitem{ku2021adaptive}
Y.-J. Ku, S.~Baidya, and S.~Dey, ``Adaptive computation partitioning and offloading in real-time sustainable vehicular edge computing,'' \emph{IEEE Transactions on Vehicular Technology}, vol.~70, no.~12, pp. 13\,221--13\,237, 2021.

\bibitem{chen2018optimized}
X.~Chen, H.~Zhang, C.~Wu, S.~Mao, Y.~Ji, and M.~Bennis, ``Optimized computation offloading performance in virtual edge computing systems via deep reinforcement learning,'' \emph{IEEE Internet of Things Journal}, vol.~6, no.~3, pp. 4005--4018, 2018.

\end{thebibliography}

\end{document}